\documentclass[aps,prl,twocolumn,superscriptaddress,floatfix,nofootinbib]{revtex4-2}
\usepackage{amsmath,amssymb,graphicx,booktabs}
\usepackage{xcolor}
\usepackage[colorlinks=true,allcolors=black]{hyperref}
\usepackage[capitalise]{cleveref}
\crefname{table}{Table}{Tables}
\crefname{figure}{Fig.}{Figs.}
\crefname{equation}{Eq.}{Eqs.}
\newcommand{\Pc}{\Phi_{\rm c}}
\newcommand{\nm}{\mathrm{n\,m^{-2}}}
\begin{document}

\title{Cold stellarator coils that outlast power plant lifetime}
\author{J. F. Parisi}
\affiliation{Marathon Fusion, 150 Mississippi Street, San Francisco, CA, 94107, USA}
\author{J. D. Suh}
\affiliation{Department of Applied Physics and Applied Mathematics, Columbia University, New York City, NY, 10027, USA}
\affiliation{Fulcrum Science, 450 Lexington Avenue, New York City, NY, 10017, USA}

\begin{abstract}
We show that a two order-of-magnitude reduction in stellarator coil peak neutron flux and heating can be achieved by moving coils much farther from the plasma if the plasma shape may change slightly and the vessel shielding is increased. Alternating coil and plasma optimizations move the closest coils of our main design over 25 cm out, without substantial degradation in key engineering and physics requirements. The field error decreases and the plasma boundary moves about 1 cm on average. Such coils would be at 4.2 K, use only one third to half the superconducting tape of the original, and require no replacement during the plant's lifetime. This reduced coil flux and heating could significantly enhance the viability of stellarator power plants.
\end{abstract}
\maketitle

\textit{Introduction.} A stellarator power plant (SPP) must make electric power for decades. But fast neutrons displace atoms in its superconducting coils, measured by the fluence above 0.1 MeV. Future SPP coils may use rare-earth barium copper oxide (REBCO) tape~\cite{wu87,molodyk21}. Design studies allow $\sim$2 to $5\cdot10^{22}$ $\nm$~\cite{sorbom15,menard16,clark25}, an uncertain limit~\cite{fischer18,adams26}. Infinity Two's average coil fluence reaches it in $\sim$4 full-power years (FPY)~\cite{clark25}. In Stellaris, the 99th-percentile coil flux reaches it in $\sim$10 FPY~\cite{lion25}. ARIES-CS and Helios coils are designed for 40 years~\cite{elguebaly08,Swanson2026}. 

These fluence limits come from fission-spectrum neutrons~\cite{torsello25}, not fusion-spectrum. Fusion-neutron tests of REBCO tape reached under a ten-thousandth of these limits (\cref{tab:dt}). Low-temperature superconductor wires reached fluences near these limits by 1983 (\cref{tab:dt}). Their room-temperature critical currents degraded significantly by $3\cdot10^{22}$ $\nm$~\cite{rtns83,guinan84}. A REBCO test at 40 K found a reversible current loss during irradiation, at far lower fluence~\cite{adams26}. Replacing modular coils means disassembling much of the machine at a high cost in hardware, lost revenue, and design complexity. Given the uncertainty in the effects of fusion neutrons, coils designed to last several SPP lifetimes ($\gtrsim$100 years) may be desirable.

Neutrons and their gamma rays also heat coils. Colder coils need more cooling, so the peak heat density limits the coil temperature. EU DEMO allows 50 W/m$^3$ in its Nb$_3$Sn toroidal field coils~\cite{fischer15,cufar21}. ARIES-CS allowed 2000 W/m$^3$~\cite{elguebaly08}. We use the stricter limit. Recent designs have REBCO near 20 K~\cite{sorbom15,lion25,Swanson2026}. At 4.2 K and about 19 T across the tape, it conducts 2 to 3 times more current~\cite{molodyk21,senatore24}. A tokamak study found REBCO coils cheaper at 4.5 K than at 20 K~\cite{chislett22}. Coils with little nuclear heat could be colder and use less tape.

More shield helps, but this needs more distance between plasma and coils. Designs leave $\sim$1.2 to 1.3 m at the closest coils for blanket and shield~\cite{elguebaly08,Swanson2026}. This least distance determines the minimum plant size~\cite{lyon94,spps97,ku08}. In ARIES-CS, 10 cm more added 60 cm of major radius~\cite{elguebaly08}. However, fine field structure needs close coils~\cite{kappel24}. The coil flux is also uneven, and the peaks determine coil life~\cite{lyytinen24}.

For tokamaks this tradeoff is well known. In ARC, 15 cm more major radius, all titanium hydride shield, lowers the coil fluence eight times~\cite{sorbom15}. In a stellarator, coils and plasma can both change shape. The closest coil parts can then move out, with the same coil length and nearly the same plasma volume.

Coil optimizations usually keep a prescribed least distance from the plasma~\cite{Wiedman2024,gil25,Swanson2026,ewers26}. For a fixed plasma, more distance has needed more conductor~\cite{kappel26,gil25}, a less accurate field~\cite{kappel24,gil25}, or reversed toroidal-field coils~\cite{henneberg26}. New winding surfaces added distance, but not at the closest points~\cite{paul18}. Single-stage and related methods adapt the plasma to its coils, but none rewards distance~\cite{strickler02,henneberg21,giuliani22,jorge23,fu26,yu26}. NCSX plasma optimization penalized sheet-current complexity~\cite{pomphrey01}, whose plasma-shape gradient was later computed at fixed coil distance~\cite{carltonjones21}. Plasma shapes have been chosen for distant coils since the 1990s~\cite{spps97,landreman16,kappel24,Goodman2024,kappel26,lion25,Swanson2026}. In ARIES-CS, coils 10\% farther out left a plasma needing improvement~\cite{ku08}. Heliotrons widened the blanket space by changing the coil pitch or adding coils~\cite{goto11,yanagi16}. ARIES-CS chose its shield thickness at the closest coils from magnet radiation limits~\cite{elguebaly08}. Elsewhere, the shield is added after the plasma or coils are fixed~\cite{moreno24}. Bogaarts and Warmer vary its thickness to limit coil flux, and propose neutronics in coil optimization~\cite{bogaarts26}.

Ku and Boozer showed that a nearby NCSX equilibrium allows simpler sheet currents 50\% farther out, at 0.21 major radii~\cite{ku10}. This Letter shows one application. With the same superconductor, coils can outlast the plant by two orders of magnitude and stay cold. Coils that only move out make a less accurate field, so we alternate two optimizations. The coil step moves the closest coil parts out radially. A small plasma step then adapts the plasma to the new coils. Our main case is QI-S3, a quasi-isodynamic reactor. Its closest coils move 27 cm out with a lower field error. Their peak flux decreases about a hundred times and fast-ion losses change minimally.

\begin{table}[tb]
\caption{Highest fluence of fusion (D-T, 14 MeV) neutrons on superconductors, to our knowledge. $T$ is temperature during irradiation, RT is `room temperature.'}
\label{tab:dt}
\begin{ruledtabular}
\begin{tabular}{lccc}
Superconductor & Fluence & $T$ & Facility, year \\
 & ($\nm$) & & \\
\hline
Nb$_3$Sn, NbTi, & 4 to $5\cdot10^{22}$ & RT & RTNS-I and II, \\
\quad V$_3$Ga wire~\cite{snead81,rtns83} & & & 1973 to 1983 \\
Nb$_3$Sn wire~\cite{guinan84} & $1.4\cdot10^{22}$ & 4 K & RTNS-II, 1983 \\
Nb$_3$Al wire~\cite{nishimura10} & $2.7\cdot10^{21}$ & RT & FNS, by 2010 \\
Bi-2223 tape~\cite{nishimura08} & $8.7\cdot10^{20}$ & & FNS, by 2008 \\
YBCO~\cite{kato88} & $2\cdot10^{19}$ & & JAERI, by 1988 \\
REBCO tape~\cite{pinto24} & $1.2\cdot10^{18}$ & RT & FNG, by 2023 \\
REBCO bridge~\cite{adams26} & $2.6\cdot10^{16}$ & 40 K & NILE, by 2025 \\
MgB$_2$, Bi-2212 & none found & & \\
\end{tabular}
\end{ruledtabular}
\end{table}

\textit{Fusion neutrons.} Fast neutrons scatter many times in the blanket and shield so their flux decreases exponentially with depth. In our model, the breeder and the other layers have the same thickness everywhere. Shield fills the rest of the space up to each coil. The point with the least shield, and so the most flux, is where a coil is closest to the plasma. Where the layers vary in thickness, the most flux is instead where the blanket plus shield is thinnest. We therefore define the coil distance $d$ as the least distance from the plasma boundary to any coil. It is measured to the center of the winding pack. The fast flux on a coil is
\begin{equation}
\Pc \simeq g\,\Phi_0\,e^{-(d-d_f)/\lambda}.
\label{eq:law}
\end{equation}
Here $\Phi_0$ is the flux behind the fixed layers alone, with no shield. The geometric factor $g$ is near one. $d_f$ is the thickness of the fixed layers: breeder, walls, vessel, gaps and half the winding pack. It is 1.20 m in QI-S3. Only the shield, of thickness $d-d_f$, increases when the coils move out. Our shield is 70\% tungsten carbide, 20\% EUROFER steel and 10\% helium by volume, at 12.5 t/m$^3$. Slab calculations with a water-cooled mix give $\lambda=5.2$ cm. The flux decreases ten times every 12 cm.

\begin{figure}[tb]
\includegraphics{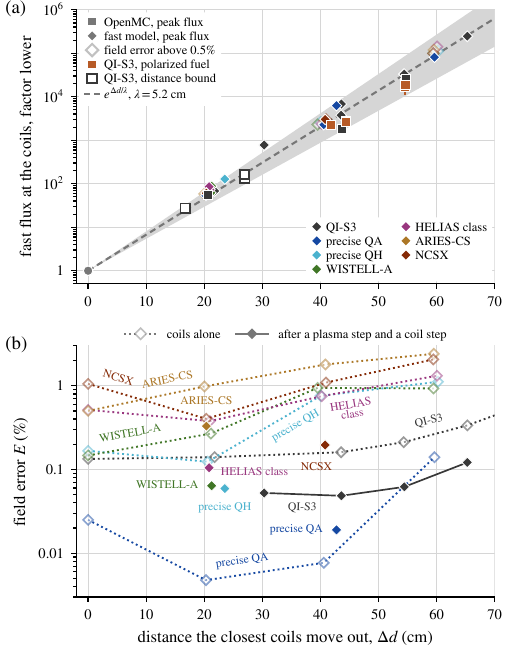}
\caption{(a) On-coil fast flux decreases exponentially with $\Delta d$. Squares are the peak flux in OpenMC. Diamonds are the peak flux in the fast model. The line is \cref{eq:law} with $\lambda=5.2$ cm, and the band has $\lambda=4.8$ to 5.9 cm. (b) Field error of the same coils: coils alone, and after one plasma step and one coil step. Here coil length is free: coils use 1.1 to 1.6 times the superconductor of the first coils.}
\label{fig:law}
\end{figure}

The plasma shape decides how far away the coils can make its field accurately. This is the main way that the plasma shape matters for the coil flux. The first wall's neutron wall load depends on the source distance and the angle of incidence~\cite{mirallesdolz26}. Behind the shield, attenuation is far more important. In our fast model, shield depth explains 99.7\% of the variation of $\ln\Phi$ along the QI-S3 coils. With no attenuation, the flux would differ by only three times between QI-S3 coil points, but just 6 cm of shield modifies the flux as much. The best that plasma shaping could do is share the flux evenly over the coils. In QI-S3, this would lower the peak about five times, but likely be a very low-performing plasma. Moving the coils out 8 cm does as much. Shaping at fixed fusion power barely changes blanket heating. The way to lower the peaks is therefore more coil distance, filled with shield.

\textit{Coil optimization.} The optimization needs the flux at every coil point, many times, so we use a fast model. For each coil point, it adds the flux from 6000 source points in the plasma along straight lines. Each line has an inverse-square factor and an attenuation from OpenMC~\cite{romano15} simulations of a slab. We then check the final coils with OpenMC and DAGMC in the full 3D geometry~\cite{wilson10}. The flux is measured in the 10 cm of each winding pack nearest the plasma. The model and OpenMC agree within a factor of two in most coil segments. At the peaks of the later coil sets, OpenMC is 1.4 to 1.8 times higher. In this work, all coil life, heating and flux values in \cref{tab:public} are from OpenMC, not the fast model.

\begin{table*}[t]
\caption{The coil sets of QI-S3 and precise QH. First: initial coils. Steps: after coil and plasma steps. Polarized: with polarized fuel. Superconductor is current times length, relative to first coils at 20 K. Life is the time for the peak flux to reach $3\cdot10^{22}$ $\nm$. Coils are at 4.2 K where the peak heat is below 50 W/m$^3$. $\Gamma_c$ and alpha loss are vacuum values. Errors are statistical.}
\label{tab:public}
\begin{ruledtabular}
\begin{tabular}{lccccc}
 & \multicolumn{3}{c}{QI-S3} & \multicolumn{2}{c}{Precise QH} \\
 & First & Steps & Polarized & First & Steps \\
\hline
Least coil distance $d$ (m) & 1.49 & 1.76 & 1.91 & 1.46 & 1.70 \\
Shield in the vessel (m$^3$) & 400 & 630 & 860 & 1030 & 1480 \\
Tritium breeding ratio & 1.22 & 1.22 & 1.23 & 1.23 & 1.23 \\
Superconductor, relative & 1 & 1.00 & 1.00 & 1 & 1.26 \\
Field error $E$ (\%) & 0.134 & 0.095 & 0.060 & 0.167 & 0.059 \\
Mean flux, factor lower & 1 & $101\pm2$ & $2670\pm90$ & 1 & $\sim$560 \\
Life (FPY) & $32\pm1$ & $4300\pm360$ & $(7.3\pm0.6)\cdot10^{4}$ & $\sim$9 & 9700 \\
Peak heat (W/m$^3$) & $112\pm5$ & $0.78\pm0.09$ & $0.030\pm0.009$ & $\sim$380 & 0.45 \\
Coil temperature (K) & 20 & 4.2 & 4.2 & 20 & 4.2 \\
Tape length, relative & 1 & 0.33 to 0.50 & 0.32 to 0.50 & 1 & 0.44 to 0.67 \\
$\Gamma_c$, relative & 1 & 0.8 to 1.1 & 0.8 to 1.1 & 1 & 2 to 11 \\
Alpha energy loss (\%) & $0.32\pm0.07$ & $0.28\pm0.06$ & $0.50\pm0.09$ & $0.052\pm0.013$ & $0.066\pm0.016$ \\
\end{tabular}
\end{ruledtabular}
\end{table*}

A coil step~\cite{Landreman2021,wechsung22} changes the coil shapes, while the coil currents and the plasma boundary stay fixed. The step minimizes $E^2$ plus penalties. Here $E$ is the field error, the root-mean-square field normal to the plasma boundary divided by the field strength. In our main case, a lower bound on the coil distance moves the closest coil parts out. Penalties keep the coil length, force, curvature and spacing within limits~\cite{hurwitz25}. A fixed coil length means a fixed amount of superconductor. We also tested adding $wF$, with a weight $w$ and the flux term
\begin{equation}
F=\frac{1}{N}\sum_{i=1}^{N}\left[\max\left(0,\ln\frac{\Phi_i}{\Phi_t}\right)\right]^2,
\label{eq:Fp}
\end{equation}
where $\Phi_i$ is the fast-model flux at point $i$ of $N$ along the coils, and $\Phi_t$ is a target. This term pushes out the coil parts with the most flux. The polarized case uses $F$.

\begin{figure}[b]
\includegraphics{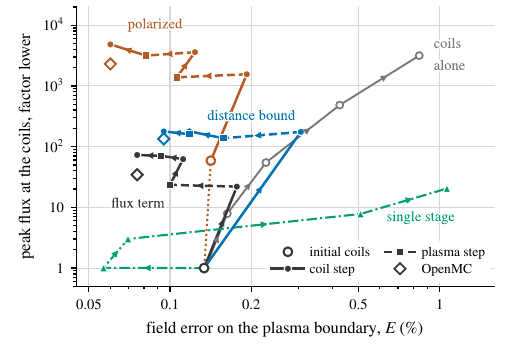}
\caption{Coil steps decrease peak flux in QI-S3 coils, and plasma steps decrease field error. Blue: steps with distance bound, our main case. Black: steps with flux term. Orange: polarized fuel. Gray: coils alone. Green: single-stage optimization. Arrows show the order of steps. Lines are from the fast model. Diamonds show the peak flux in OpenMC.}
\label{fig:loop}
\end{figure}

A plasma step then changes the plasma boundary with the coils fixed, using the DESC code~\cite{dudt20,panici23,dudt23}. It minimizes $E$ and penalizes changes of the plasma volume. It also limits $\Gamma_c$, a proxy for fast-ion confinement~\cite{nemov08,velasco21}. This proxy averages the radial drift of trapped fast ions relative to their drift around the flux surface. Coils moved out on the first plasma, with no plasma step, show how much error the alphas tolerate. At a field error of 0.25\%, as many alphas are lost as with the first coils. At 0.47\% twice as many are lost, and at 0.71\% eight times as many. How much error a plasma tolerates depends on the error's pattern. Ku and Boozer found NCSX most sensitive to resonant patterns and to errors on the outboard side~\cite{ku10}. As a conservative rule, we require the final coils never exceed the error of the first coils.

\begin{figure*}[t]
\includegraphics{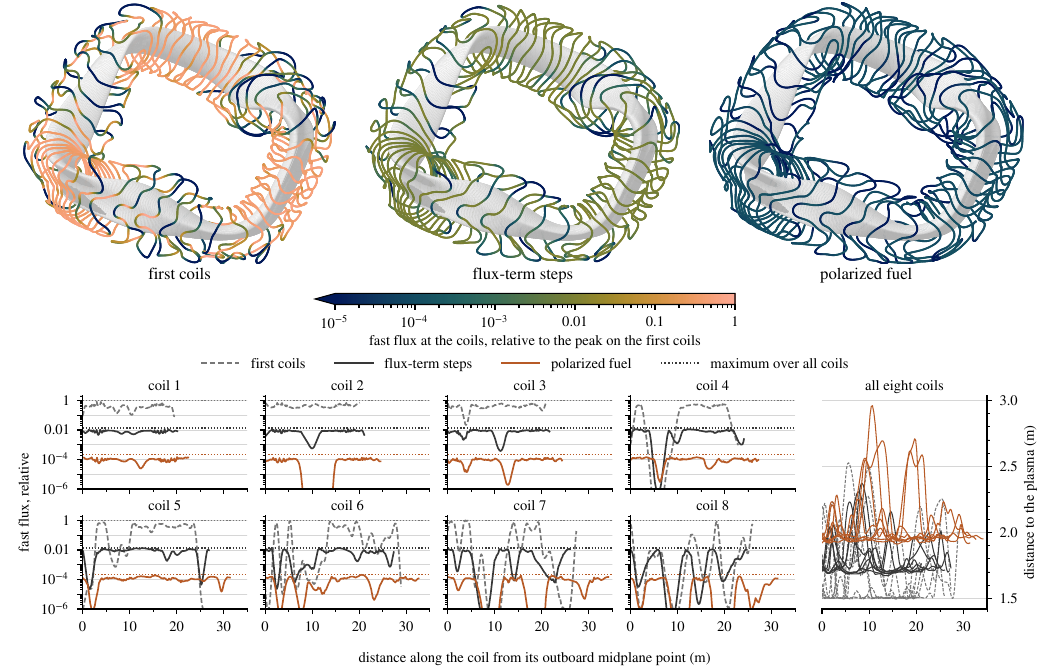}
\caption{The QI-S3 coils, colored by the fast flux of the fast model, relative to the peak on the first coils. Shown are steps with an earlier flux term. They move the closest coils 18 cm out, or 42 cm with polarized fuel. Below, the same flux along each of the eight distinct coils, and at right their distance to the plasma.}
\label{fig:coils}
\end{figure*}

With a fixed plasma and the same amount of superconductor, coils alone cannot move far. In a slab, the field of a coil feature of size $\ell$ decreases as $e^{-2\pi d/\ell}$ with distance $d$~\cite{landreman16,ku10}. In QI-S3, coils moved out 22 cm alone increase the error from 0.134\% to 0.23\% (gray in \cref{fig:loop}). Their field lines still form good surfaces, so such coils may be acceptable. But they break our conservative rule, since their error exceeds that of the first coils. The plasma step removes this limit by changing the plasma slightly. The boundary moves about 1 cm on average and at most 6 cm. The harmonics of its field strength change by under 4\% up to order 7. What changes is the fine structure that the coils must make. An ideal coil, a current sheet 1.7 m from the plasma, needs about three times less current above order 5. For a fixed plasma, farther coils are more complex at the same field error~\cite{kappel24}. With the steps, an ideal sheet needs 15\% more peak current, versus 2.5 times on the first plasma. More superconductor also allows distance~\cite{wechsung22}: with 1.4 times as much, coils alone reach 44 cm at $E=0.16\%$ (\cref{fig:law}b).

\textit{Results.} QI-S3 is a vacuum equilibrium with three field periods. Its major radius $R$ is 10.1 m and its minor radius $a$ is 1.09 m. A zero-dimensional model with assumed density and temperature profiles produces 2.4 GW of fusion power. This needs a mean plasma pressure 3.5\% of the magnetic pressure. We did not solve the equilibrium at this pressure. The mean neutron wall load is 2.9 MW/m$^2$. Its 48 first coils are $1.37a$ from the plasma. Three coil steps (with a distance bound) and two plasma steps move the closest coils 27 cm out. The amount of superconductor is unchanged and the field error decreases from 0.134\% to 0.095\%.

In OpenMC, the peak fast flux on the coils decreases $134\pm13$ times. The coil life increases from 32 to $4300\pm360$ FPY. We fill the space around the coils with shield, because neutrons otherwise reach their sides and backs. As modeled, this fill is about 7000 m$^3$, far more than a design would use. Without the fill, both lifetimes are three to four times shorter. The first coils, with this idealized shield, are already near both limits. A few cm more shield would meet both. The steps instead give a margin of two orders of magnitude.

$\Gamma_c$ and the alpha loss change within their errors. We also followed 4000 alphas for 0.1 s in each coil field. They lose $0.18\pm0.07\%$ of their energy with the first coils and $0.25\pm0.07\%$ after the steps. Traced field lines show no broken surfaces inside the plasma with the final coils.

The added distance has a tradeoff. In QI-S3, the plasma volume decreases by 0.2\%. The shield in the vessel increases from 400 to 630 m$^3$, or from 5000 to 7900 t. Its thinnest place increases from 30 to 57 cm. Tritium breeding does not change, with a ratio of 1.22 at full blanket coverage.

If we use the flux term instead of the bound, the same steps move the coils out 21 cm. The peak flux then decreases $54\pm4$ times. Our single-stage optimizations, started from the first coils, did worse: at similar distances, their field error was ten to twenty times higher (green in \cref{fig:loop}). With 1.17 times the superconductor, coils on the stepped plasma move 44 cm out, with an error of 0.053\%. Their peak flux then decreases $1800\pm400$ times in OpenMC, and their alpha loss changes within errors. Polarized fuel lets the coils move farther out. Aligned spins increase the D-T cross section by up to 1.5 times~\cite{kulsrud82,parisi24,bae25}. The same fusion power then needs less field~\cite{pace16,baylor23}. If confinement limits QI-S3, it needs 16\% less field so its coils can start 22 cm farther out. The steps move them 42 cm in total, and the peak flux decreases by a factor of $2300\pm200$ (\cref{fig:coils}). This includes the neutrons that aligned spins send across the field~\cite{kulsrud82,bae25}. The alpha energy loss increases from 0.32\% to 0.50\%. The fuel causes this, not the steps: the first plasma with polarized fuel also loses 0.50\%.

\textit{Other stellarators.} QI-S3 was not designed for stability, but W7-X keeps its Mercier stability through the steps. QI-S3 was shaped for fast-ion confinement and coil distance, so it only tests the distance. It is Mercier unstable from $\beta=0.25\%$, before and after the steps, where $\beta$ is the ratio of plasma pressure to magnetic pressure. For stability, we use W7-X at $\beta=4.5\%$~\cite{Grieger1992}. One coil step, one plasma step and a second coil step move its closest coils 21 cm out. Its field error ends at 0.31\%, below the 0.51\% of its first coils. We solved its fixed-boundary equilibrium before and after the steps with VMEC++~\cite{Schilling2025}. From $\beta=3\%$ to 5.5\%, both are Mercier stable outside the inner quarter of the radius.

Precise QH, a published quasi-helical vacuum equilibrium with four periods~\cite{landreman22}, is a harder case. With almost the same coil length, one coil step moves its closest coils only 4 cm out. With 1.26 times the superconductor, the steps move them 24 cm out (\cref{tab:public}). Its mean flux then decreases several hundred times, it loses 1.8\% of its plasma volume, and its shield increases from 1030 to 1480 m$^3$. Its peak force also increases, and its $\Gamma_c$ increases 2 to 11 times, to about the value in QI-S3. 

We measure the coil distance an equilibrium allows with $d_*$. It is the farthest distance at which a sheet of current can still make the field~\cite{landreman17}. We use the limits of Kappel \textit{et al.}: $E=0.17\%$ and 17 MA/m at 5.9 T~\cite{kappel24}. In 98 equilibria, 12 published, 10 from the ConStellaration dataset~\cite{Cadena2025} and 76 that we optimized from these, we find
\begin{equation}
\frac{d_*}{a}\simeq0.68\left(\frac{A}{N}\right)^{0.58},
\label{eq:dstar}
\end{equation}
with a standard deviation of 19\% about the fit. Here $A=R/a$ is the aspect ratio and $N$ the number of field periods. Equilibria with few periods and high aspect ratio thus allow the most distance, in units of $a$. This agrees with Kappel \textit{et al.}, who found that few field periods allow more distance~\cite{kappel24}. Earlier work found simpler coils at larger $A/N$, without a fitted law~\cite{kuboozer10nf,Cadena2025,kappel26}. We also apply the steps to six published equilibria (\cref{fig:law}b). One plasma and one coil step move their coils 20 to 43 cm out, with 10 to 30\% more superconductor. For three of them the error is 3 to 4 times lower than with coils alone. They are a HELIAS-class equilibrium~\cite{Grieger1992}, WISTELL-A~\cite{bader20} and precise QH. Precise QA needs no plasma step. ARIES-CS~\cite{najmabadi08,ku08} and NCSX~\cite{Zarnstorff2001}, at aspect ratios near 4.5, leave the least room (\cref{eq:dstar}).

\textit{Discussion.} We have shown that optimizing for lower coil flux and heating significantly opens the stellarator design space, reducing coil flux and heating by orders of magnitude. This surprising decrease in flux and heating is due to the exponential flux and heat decrease with shielding. There is minimal decrease in plasma parameters such as fast-ion confinement and field error, consistent with Ku and Boozer's findings that coils can be located much farther from the plasma boundary~\cite{ku10}. We also required that the optimization does not require unreasonably complex coil manufacturing, by requiring the peak forces and tape strains remain reasonable. By allowing coils to be much farther from the plasma, this loosens a very strong constraint on stellarator power plant design~\cite{lyon94,spps97}, and could allow for stellarators with improved economics. The increased distance could also ease other constraints of a plant. It could contain a thicker breeding blanket, or space for other subsystems. Stellarators with more space may also generate extra value, such as with mercury blankets that produce 1-2 tons of gold per gigawatt thermal year~\cite{Rutkowski2025}.

A rough estimate shows the economic benefit of avoiding coil replacement. At \$150 to 200 per kA m~\cite{molodyk23}, the tape of the first QI-S3 coils costs about \$2 billion. Coils at 4.2 K only need a third to half of it. Coils that last a decade need three replacement sets in 40 FPY, $\sim$\$6 billion of tape. Each year of outage also loses about \$0.5 billion in electricity sales, for 1 GW$_\mathrm{e}$ at \$60 per MWh$_\mathrm{e}$. These figures exclude the larger shield and the added cooling.

This is an initial study with many caveats and avenues for future work. Much more work is needed to fully characterize the tradeoffs with coil-plasma distance. A non-exhaustive list is neoclassical~\cite{nemov99,beidler21} and turbulent transport~\cite{mynick10,Goodman2024}, global instabilities~\cite{schwab93}, first-wall neutron loading~\cite{lion22,mirallesdolz26}, cooling, cost and waste of the larger shield~\cite{elguebaly08}, and non-uniform blanket and shielding build~\cite{lion25,mirallesdolz26sr}. Spin-polarized fuel~\cite{kulsrud82,parisi24} is speculative until it is tested in magnetic confinement~\cite{heidbrink24,baylor23}.

Over 40 FPY, such coils receive about 1\% of the fluence limit. Irradiation tests of new tapes could then stop at a far lower fluence, which would speed their qualification. Colder coils could also use other superconductor materials than REBCO~\cite{risse17,sarasola20}, or, where stresses allow, produce a higher field.

In summary, moving coils farther away would significantly reduce the neutron flux and heating on superconductors in stellarator coils. This could significantly reduce the probability of the coils failing during operation, as well as opening the design space for stellarator power plants with improved economics. These results motivate a fuller study of coil distance and its tradeoffs in a stellarator power plant.

\textit{Acknowledgments.} We are grateful for discussions with A. Diallo, E. J. Paul, and A. Rutkowski.

\textit{AI Statement and Data Availability.} We made extensive use of Claude Opus 5.5 and Claude Fable 5.2 in all aspects of this project. All scripts required to reproduce results of this paper will be posted in a public, permanent repository.

\bibliography{letter}

\begin{thebibliography}{84}%
\makeatletter
\providecommand \@ifxundefined [1]{%
 \@ifx{#1\undefined}
}%
\providecommand \@ifnum [1]{%
 \ifnum #1\expandafter \@firstoftwo
 \else \expandafter \@secondoftwo
 \fi
}%
\providecommand \@ifx [1]{%
 \ifx #1\expandafter \@firstoftwo
 \else \expandafter \@secondoftwo
 \fi
}%
\providecommand \natexlab [1]{#1}%
\providecommand \enquote  [1]{``#1''}%
\providecommand \bibnamefont  [1]{#1}%
\providecommand \bibfnamefont [1]{#1}%
\providecommand \citenamefont [1]{#1}%
\providecommand \href@noop [0]{\@secondoftwo}%
\providecommand \href [0]{\begingroup \@sanitize@url \@href}%
\providecommand \@href[1]{\@@startlink{#1}\@@href}%
\providecommand \@@href[1]{\endgroup#1\@@endlink}%
\providecommand \@sanitize@url [0]{\catcode `\\12\catcode `\$12\catcode
  `\&12\catcode `\#12\catcode `\^12\catcode `\_12\catcode `\%12\relax}%
\providecommand \@@startlink[1]{}%
\providecommand \@@endlink[0]{}%
\providecommand \url  [0]{\begingroup\@sanitize@url \@url }%
\providecommand \@url [1]{\endgroup\@href {#1}{\urlprefix }}%
\providecommand \urlprefix  [0]{URL }%
\providecommand \Eprint [0]{\href }%
\providecommand \doibase [0]{https://doi.org/}%
\providecommand \selectlanguage [0]{\@gobble}%
\providecommand \bibinfo  [0]{\@secondoftwo}%
\providecommand \bibfield  [0]{\@secondoftwo}%
\providecommand \translation [1]{[#1]}%
\providecommand \BibitemOpen [0]{}%
\providecommand \bibitemStop [0]{}%
\providecommand \bibitemNoStop [0]{.\EOS\space}%
\providecommand \EOS [0]{\spacefactor3000\relax}%
\providecommand \BibitemShut  [1]{\csname bibitem#1\endcsname}%
\let\auto@bib@innerbib\@empty
\bibitem [{\citenamefont {Wu}\ \emph {et~al.}(1987)\citenamefont {Wu},
  \citenamefont {Ashburn}, \citenamefont {Torng}, \citenamefont {Hor},
  \citenamefont {Meng}, \citenamefont {Gao}, \citenamefont {Huang},
  \citenamefont {Wang},\ and\ \citenamefont {Chu}}]{wu87}%
  \BibitemOpen
  \bibfield  {author} {\bibinfo {author} {\bibfnamefont {M.~K.}\ \bibnamefont
  {Wu}}, \bibinfo {author} {\bibfnamefont {J.~R.}\ \bibnamefont {Ashburn}},
  \bibinfo {author} {\bibfnamefont {C.~J.}\ \bibnamefont {Torng}}, \bibinfo
  {author} {\bibfnamefont {P.~H.}\ \bibnamefont {Hor}}, \bibinfo {author}
  {\bibfnamefont {R.~L.}\ \bibnamefont {Meng}}, \bibinfo {author}
  {\bibfnamefont {L.}~\bibnamefont {Gao}}, \bibinfo {author} {\bibfnamefont
  {Z.~J.}\ \bibnamefont {Huang}}, \bibinfo {author} {\bibfnamefont {Y.~Q.}\
  \bibnamefont {Wang}},\ and\ \bibinfo {author} {\bibfnamefont {C.~W.}\
  \bibnamefont {Chu}},\ }\bibfield  {title} {\bibinfo {title}
  {Superconductivity at 93 {K} in a new mixed-phase {Y-Ba-Cu-O} compound system
  at ambient pressure},\ }\href {https://doi.org/10.1103/PhysRevLett.58.908}
  {\bibfield  {journal} {\bibinfo  {journal} {Phys. Rev. Lett.}\ }\textbf
  {\bibinfo {volume} {58}},\ \bibinfo {pages} {908} (\bibinfo {year}
  {1987})}\BibitemShut {NoStop}%
\bibitem [{\citenamefont {Molodyk}\ \emph {et~al.}(2021)\citenamefont
  {Molodyk}, \citenamefont {Samoilenkov}, \citenamefont {Markelov},
  \citenamefont {Degtyarenko}, \citenamefont {Lee}, \citenamefont {Petrykin},
  \citenamefont {Gaifullin}, \citenamefont {Mankevich}, \citenamefont
  {Vavilov}, \citenamefont {Sorbom} \emph {et~al.}}]{molodyk21}%
  \BibitemOpen
  \bibfield  {author} {\bibinfo {author} {\bibfnamefont {A.}~\bibnamefont
  {Molodyk}}, \bibinfo {author} {\bibfnamefont {S.}~\bibnamefont
  {Samoilenkov}}, \bibinfo {author} {\bibfnamefont {A.}~\bibnamefont
  {Markelov}}, \bibinfo {author} {\bibfnamefont {P.}~\bibnamefont
  {Degtyarenko}}, \bibinfo {author} {\bibfnamefont {S.}~\bibnamefont {Lee}},
  \bibinfo {author} {\bibfnamefont {V.}~\bibnamefont {Petrykin}}, \bibinfo
  {author} {\bibfnamefont {M.}~\bibnamefont {Gaifullin}}, \bibinfo {author}
  {\bibfnamefont {A.}~\bibnamefont {Mankevich}}, \bibinfo {author}
  {\bibfnamefont {A.}~\bibnamefont {Vavilov}}, \bibinfo {author} {\bibfnamefont
  {B.}~\bibnamefont {Sorbom}}, \emph {et~al.},\ }\bibfield  {title} {\bibinfo
  {title} {Development and large volume production of extremely high current
  density {YBa$_2$Cu$_3$O$_7$} superconducting wires for fusion},\ }\href
  {https://doi.org/10.1038/s41598-021-81559-z} {\bibfield  {journal} {\bibinfo
  {journal} {Sci. Rep.}\ }\textbf {\bibinfo {volume} {11}},\ \bibinfo {pages}
  {2084} (\bibinfo {year} {2021})}\BibitemShut {NoStop}%
\bibitem [{\citenamefont {Sorbom}\ \emph {et~al.}(2015)\citenamefont {Sorbom},
  \citenamefont {Ball}, \citenamefont {Palmer}, \citenamefont {Mangiarotti},
  \citenamefont {Sierchio}, \citenamefont {Bonoli}, \citenamefont {Kasten},
  \citenamefont {Sutherland}, \citenamefont {Barnard}, \citenamefont
  {Haakonsen} \emph {et~al.}}]{sorbom15}%
  \BibitemOpen
  \bibfield  {author} {\bibinfo {author} {\bibfnamefont {B.~N.}\ \bibnamefont
  {Sorbom}}, \bibinfo {author} {\bibfnamefont {J.}~\bibnamefont {Ball}},
  \bibinfo {author} {\bibfnamefont {T.~R.}\ \bibnamefont {Palmer}}, \bibinfo
  {author} {\bibfnamefont {F.~J.}\ \bibnamefont {Mangiarotti}}, \bibinfo
  {author} {\bibfnamefont {J.~M.}\ \bibnamefont {Sierchio}}, \bibinfo {author}
  {\bibfnamefont {P.}~\bibnamefont {Bonoli}}, \bibinfo {author} {\bibfnamefont
  {C.}~\bibnamefont {Kasten}}, \bibinfo {author} {\bibfnamefont {D.~A.}\
  \bibnamefont {Sutherland}}, \bibinfo {author} {\bibfnamefont {H.~S.}\
  \bibnamefont {Barnard}}, \bibinfo {author} {\bibfnamefont {C.~B.}\
  \bibnamefont {Haakonsen}}, \emph {et~al.},\ }\bibfield  {title} {\bibinfo
  {title} {{ARC}: {A} compact, high-field, fusion nuclear science facility and
  demonstration power plant with demountable magnets},\ }\href
  {https://doi.org/10.1016/j.fusengdes.2015.07.008} {\bibfield  {journal}
  {\bibinfo  {journal} {Fusion Eng. Des.}\ }\textbf {\bibinfo {volume} {100}},\
  \bibinfo {pages} {378} (\bibinfo {year} {2015})}\BibitemShut {NoStop}%
\bibitem [{\citenamefont {Menard}\ \emph {et~al.}(2016)\citenamefont {Menard},
  \citenamefont {Brown}, \citenamefont {El-Guebaly}, \citenamefont {Boyer},
  \citenamefont {Canik}, \citenamefont {Colling}, \citenamefont {Raman},
  \citenamefont {Wang}, \citenamefont {Zhai}, \citenamefont {Buxton} \emph
  {et~al.}}]{menard16}%
  \BibitemOpen
  \bibfield  {author} {\bibinfo {author} {\bibfnamefont {J.~E.}\ \bibnamefont
  {Menard}}, \bibinfo {author} {\bibfnamefont {T.}~\bibnamefont {Brown}},
  \bibinfo {author} {\bibfnamefont {L.}~\bibnamefont {El-Guebaly}}, \bibinfo
  {author} {\bibfnamefont {M.}~\bibnamefont {Boyer}}, \bibinfo {author}
  {\bibfnamefont {J.}~\bibnamefont {Canik}}, \bibinfo {author} {\bibfnamefont
  {B.}~\bibnamefont {Colling}}, \bibinfo {author} {\bibfnamefont
  {R.}~\bibnamefont {Raman}}, \bibinfo {author} {\bibfnamefont
  {Z.}~\bibnamefont {Wang}}, \bibinfo {author} {\bibfnamefont {Y.}~\bibnamefont
  {Zhai}}, \bibinfo {author} {\bibfnamefont {P.}~\bibnamefont {Buxton}}, \emph
  {et~al.},\ }\bibfield  {title} {\bibinfo {title} {Fusion nuclear science
  facilities and pilot plants based on the spherical tokamak},\ }\href
  {https://doi.org/10.1088/0029-5515/56/10/106023} {\bibfield  {journal}
  {\bibinfo  {journal} {Nucl. Fusion}\ }\textbf {\bibinfo {volume} {56}},\
  \bibinfo {pages} {106023} (\bibinfo {year} {2016})}\BibitemShut {NoStop}%
\bibitem [{\citenamefont {Clark}\ \emph {et~al.}(2025)\citenamefont {Clark},
  \citenamefont {Goh}, \citenamefont {Ramirez}, \citenamefont {Pflug},
  \citenamefont {Smandych}, \citenamefont {Kessing}, \citenamefont {Moreno},
  \citenamefont {Bohm}, \citenamefont {Wilson}, \citenamefont {Singh} \emph
  {et~al.}}]{clark25}%
  \BibitemOpen
  \bibfield  {author} {\bibinfo {author} {\bibfnamefont {D.~W.~S.}\
  \bibnamefont {Clark}}, \bibinfo {author} {\bibfnamefont {B.}~\bibnamefont
  {Goh}}, \bibinfo {author} {\bibfnamefont {S.}~\bibnamefont {Ramirez}},
  \bibinfo {author} {\bibfnamefont {E.}~\bibnamefont {Pflug}}, \bibinfo
  {author} {\bibfnamefont {J.}~\bibnamefont {Smandych}}, \bibinfo {author}
  {\bibfnamefont {J.~R.}\ \bibnamefont {Kessing}}, \bibinfo {author}
  {\bibfnamefont {C.}~\bibnamefont {Moreno}}, \bibinfo {author} {\bibfnamefont
  {T.~D.}\ \bibnamefont {Bohm}}, \bibinfo {author} {\bibfnamefont {P.~P.~H.}\
  \bibnamefont {Wilson}}, \bibinfo {author} {\bibfnamefont {L.}~\bibnamefont
  {Singh}}, \emph {et~al.},\ }\bibfield  {title} {\bibinfo {title} {Breeder
  blanket and tritium fuel cycle feasibility of the {Infinity Two} fusion pilot
  plant},\ }\href {https://doi.org/10.1017/S002237782500039X} {\bibfield
  {journal} {\bibinfo  {journal} {J. Plasma Phys.}\ }\textbf {\bibinfo {volume}
  {91}},\ \bibinfo {pages} {E86} (\bibinfo {year} {2025})}\BibitemShut
  {NoStop}%
\bibitem [{\citenamefont {Fischer}\ \emph {et~al.}(2018)\citenamefont
  {Fischer}, \citenamefont {Prokopec}, \citenamefont {Emhofer},\ and\
  \citenamefont {Eisterer}}]{fischer18}%
  \BibitemOpen
  \bibfield  {author} {\bibinfo {author} {\bibfnamefont {D.~X.}\ \bibnamefont
  {Fischer}}, \bibinfo {author} {\bibfnamefont {R.}~\bibnamefont {Prokopec}},
  \bibinfo {author} {\bibfnamefont {J.}~\bibnamefont {Emhofer}},\ and\ \bibinfo
  {author} {\bibfnamefont {M.}~\bibnamefont {Eisterer}},\ }\bibfield  {title}
  {\bibinfo {title} {The effect of fast neutron irradiation on the
  superconducting properties of {REBCO} coated conductors with and without
  artificial pinning centers},\ }\href
  {https://doi.org/10.1088/1361-6668/aaadf2} {\bibfield  {journal} {\bibinfo
  {journal} {Supercond. Sci. Technol.}\ }\textbf {\bibinfo {volume} {31}},\
  \bibinfo {pages} {044006} (\bibinfo {year} {2018})}\BibitemShut {NoStop}%
\bibitem [{\citenamefont {Adams}\ \emph {et~al.}(2026)\citenamefont {Adams},
  \citenamefont {Winger}, \citenamefont {Lewis}, \citenamefont {Cazzaniga},
  \citenamefont {Frost}, \citenamefont {Kastriotou}, \citenamefont {Iliffe},
  \citenamefont {Grovenor},\ and\ \citenamefont {Speller}}]{adams26}%
  \BibitemOpen
  \bibfield  {author} {\bibinfo {author} {\bibfnamefont {K.}~\bibnamefont
  {Adams}}, \bibinfo {author} {\bibfnamefont {J.~R.}\ \bibnamefont {Winger}},
  \bibinfo {author} {\bibfnamefont {J.~C.}\ \bibnamefont {Lewis}}, \bibinfo
  {author} {\bibfnamefont {C.}~\bibnamefont {Cazzaniga}}, \bibinfo {author}
  {\bibfnamefont {C.}~\bibnamefont {Frost}}, \bibinfo {author} {\bibfnamefont
  {M.}~\bibnamefont {Kastriotou}}, \bibinfo {author} {\bibfnamefont
  {W.}~\bibnamefont {Iliffe}}, \bibinfo {author} {\bibfnamefont {C.~R.~M.}\
  \bibnamefont {Grovenor}},\ and\ \bibinfo {author} {\bibfnamefont {S.~C.}\
  \bibnamefont {Speller}},\ }\bibfield  {title} {\bibinfo {title} {The
  performance of {REBCO} coated conductor during in situ cryogenic irradiation
  with fusion-spectrum neutrons},\ }\href
  {https://doi.org/10.1088/1361-6668/ae4548} {\bibfield  {journal} {\bibinfo
  {journal} {Supercond. Sci. Technol.}\ }\textbf {\bibinfo {volume} {39}},\
  \bibinfo {pages} {02LT01} (\bibinfo {year} {2026})}\BibitemShut {NoStop}%
\bibitem [{\citenamefont {Lion}\ \emph {et~al.}(2025)\citenamefont {Lion},
  \citenamefont {Angl{\`e}s}, \citenamefont {Bonauer}, \citenamefont
  {Ba{\~n}{\'o}n~Navarro}, \citenamefont {Cadena~Ceron}, \citenamefont
  {Davies}, \citenamefont {Drevlak}, \citenamefont {Foppiani}, \citenamefont
  {Geiger}, \citenamefont {Goodman} \emph {et~al.}}]{lion25}%
  \BibitemOpen
  \bibfield  {author} {\bibinfo {author} {\bibfnamefont {J.}~\bibnamefont
  {Lion}}, \bibinfo {author} {\bibfnamefont {J.-C.}\ \bibnamefont
  {Angl{\`e}s}}, \bibinfo {author} {\bibfnamefont {L.}~\bibnamefont {Bonauer}},
  \bibinfo {author} {\bibfnamefont {A.}~\bibnamefont {Ba{\~n}{\'o}n~Navarro}},
  \bibinfo {author} {\bibfnamefont {S.~A.}\ \bibnamefont {Cadena~Ceron}},
  \bibinfo {author} {\bibfnamefont {R.}~\bibnamefont {Davies}}, \bibinfo
  {author} {\bibfnamefont {M.}~\bibnamefont {Drevlak}}, \bibinfo {author}
  {\bibfnamefont {N.}~\bibnamefont {Foppiani}}, \bibinfo {author}
  {\bibfnamefont {J.}~\bibnamefont {Geiger}}, \bibinfo {author} {\bibfnamefont
  {A.}~\bibnamefont {Goodman}}, \emph {et~al.},\ }\bibfield  {title} {\bibinfo
  {title} {{Stellaris}: {A} high-field quasi-isodynamic stellarator for a
  prototypical fusion power plant},\ }\href
  {https://doi.org/10.1016/j.fusengdes.2025.114868} {\bibfield  {journal}
  {\bibinfo  {journal} {Fusion Eng. Des.}\ }\textbf {\bibinfo {volume} {214}},\
  \bibinfo {pages} {114868} (\bibinfo {year} {2025})}\BibitemShut {NoStop}%
\bibitem [{\citenamefont {El-Guebaly}\ \emph {et~al.}(2008)\citenamefont
  {El-Guebaly}, \citenamefont {Wilson}, \citenamefont {Henderson},
  \citenamefont {Sawan}, \citenamefont {Sviatoslavsky}, \citenamefont
  {Tautges}, \citenamefont {Slaybaugh}, \citenamefont {Kiedrowski},
  \citenamefont {Ibrahim}, \citenamefont {Martin} \emph
  {et~al.}}]{elguebaly08}%
  \BibitemOpen
  \bibfield  {author} {\bibinfo {author} {\bibfnamefont {L.}~\bibnamefont
  {El-Guebaly}}, \bibinfo {author} {\bibfnamefont {P.}~\bibnamefont {Wilson}},
  \bibinfo {author} {\bibfnamefont {D.}~\bibnamefont {Henderson}}, \bibinfo
  {author} {\bibfnamefont {M.}~\bibnamefont {Sawan}}, \bibinfo {author}
  {\bibfnamefont {G.}~\bibnamefont {Sviatoslavsky}}, \bibinfo {author}
  {\bibfnamefont {T.}~\bibnamefont {Tautges}}, \bibinfo {author} {\bibfnamefont
  {R.}~\bibnamefont {Slaybaugh}}, \bibinfo {author} {\bibfnamefont
  {B.}~\bibnamefont {Kiedrowski}}, \bibinfo {author} {\bibfnamefont
  {A.}~\bibnamefont {Ibrahim}}, \bibinfo {author} {\bibfnamefont
  {C.}~\bibnamefont {Martin}}, \emph {et~al.},\ }\bibfield  {title} {\bibinfo
  {title} {Designing {ARIES-CS} compact radial build and nuclear system:
  Neutronics, shielding, and activation},\ }\href
  {https://doi.org/10.13182/FST54-747} {\bibfield  {journal} {\bibinfo
  {journal} {Fusion Sci. Technol.}\ }\textbf {\bibinfo {volume} {54}},\
  \bibinfo {pages} {747} (\bibinfo {year} {2008})}\BibitemShut {NoStop}%
\bibitem [{\citenamefont {Swanson}\ \emph {et~al.}(2026)\citenamefont
  {Swanson}, \citenamefont {Kumar}, \citenamefont {Dudt}, \citenamefont {Flom},
  \citenamefont {Kalb}, \citenamefont {Kruger}, \citenamefont {Martin},
  \citenamefont {Olatunji}, \citenamefont {Pasmann}, \citenamefont {Tang},
  \citenamefont {von~der Linden}, \citenamefont {Wasserman}, \citenamefont
  {Avida}, \citenamefont {Basurto}, \citenamefont {Dickerson}, \citenamefont
  {de~Boer}, \citenamefont {Donovan}, \citenamefont {Doudna~Cate},
  \citenamefont {Fort}, \citenamefont {Harris}, \citenamefont {Khera},
  \citenamefont {Koen}, \citenamefont {Labbate}, \citenamefont {Maitra},
  \citenamefont {Ottaviano}, \citenamefont {Parmar}, \citenamefont {Paul},
  \citenamefont {Reydel}, \citenamefont {van Riel}, \citenamefont {Romano},
  \citenamefont {Savastianov}, \citenamefont {Saxena}, \citenamefont
  {Seethalla}, \citenamefont {Srinivasan}, \citenamefont {Wu}, \citenamefont
  {Nash}, \citenamefont {Priebe}, \citenamefont {Slepchenkov}, \citenamefont
  {Walsh}, \citenamefont {Berzin},\ and\ \citenamefont {Gates}}]{Swanson2026}%
  \BibitemOpen
  \bibfield  {author} {\bibinfo {author} {\bibfnamefont {C.~P.~S.}\
  \bibnamefont {Swanson}}, \bibinfo {author} {\bibfnamefont {S.~T.~A.}\
  \bibnamefont {Kumar}}, \bibinfo {author} {\bibfnamefont {D.~W.}\ \bibnamefont
  {Dudt}}, \bibinfo {author} {\bibfnamefont {E.~R.}\ \bibnamefont {Flom}},
  \bibinfo {author} {\bibfnamefont {W.~B.}\ \bibnamefont {Kalb}}, \bibinfo
  {author} {\bibfnamefont {T.~G.}\ \bibnamefont {Kruger}}, \bibinfo {author}
  {\bibfnamefont {M.~F.}\ \bibnamefont {Martin}}, \bibinfo {author}
  {\bibfnamefont {J.~R.}\ \bibnamefont {Olatunji}}, \bibinfo {author}
  {\bibfnamefont {S.}~\bibnamefont {Pasmann}}, \bibinfo {author} {\bibfnamefont
  {L.~Z.}\ \bibnamefont {Tang}}, \bibinfo {author} {\bibfnamefont
  {J.}~\bibnamefont {von~der Linden}}, \bibinfo {author} {\bibfnamefont
  {J.}~\bibnamefont {Wasserman}}, \bibinfo {author} {\bibfnamefont
  {M.}~\bibnamefont {Avida}}, \bibinfo {author} {\bibfnamefont {A.~S.}\
  \bibnamefont {Basurto}}, \bibinfo {author} {\bibfnamefont {M.}~\bibnamefont
  {Dickerson}}, \bibinfo {author} {\bibfnamefont {N.}~\bibnamefont {de~Boer}},
  \bibinfo {author} {\bibfnamefont {M.~J.}\ \bibnamefont {Donovan}}, \bibinfo
  {author} {\bibfnamefont {A.~H.}\ \bibnamefont {Doudna~Cate}}, \bibinfo
  {author} {\bibfnamefont {D.}~\bibnamefont {Fort}}, \bibinfo {author}
  {\bibfnamefont {W.}~\bibnamefont {Harris}}, \bibinfo {author} {\bibfnamefont
  {U.}~\bibnamefont {Khera}}, \bibinfo {author} {\bibfnamefont
  {A.}~\bibnamefont {Koen}}, \bibinfo {author} {\bibfnamefont {J.~A.}\
  \bibnamefont {Labbate}}, \bibinfo {author} {\bibfnamefont {N.}~\bibnamefont
  {Maitra}}, \bibinfo {author} {\bibfnamefont {A.}~\bibnamefont {Ottaviano}},
  \bibinfo {author} {\bibfnamefont {R.~K.}\ \bibnamefont {Parmar}}, \bibinfo
  {author} {\bibfnamefont {E.~J.}\ \bibnamefont {Paul}}, \bibinfo {author}
  {\bibfnamefont {B.}~\bibnamefont {Reydel}}, \bibinfo {author} {\bibfnamefont
  {A.}~\bibnamefont {van Riel}}, \bibinfo {author} {\bibfnamefont {P.~K.}\
  \bibnamefont {Romano}}, \bibinfo {author} {\bibfnamefont {M.}~\bibnamefont
  {Savastianov}}, \bibinfo {author} {\bibfnamefont {S.}~\bibnamefont {Saxena}},
  \bibinfo {author} {\bibfnamefont {S.}~\bibnamefont {Seethalla}}, \bibinfo
  {author} {\bibfnamefont {S.}~\bibnamefont {Srinivasan}}, \bibinfo {author}
  {\bibfnamefont {R.~H.}\ \bibnamefont {Wu}}, \bibinfo {author} {\bibfnamefont
  {D.}~\bibnamefont {Nash}}, \bibinfo {author} {\bibfnamefont {J.}~\bibnamefont
  {Priebe}}, \bibinfo {author} {\bibfnamefont {M.}~\bibnamefont {Slepchenkov}},
  \bibinfo {author} {\bibfnamefont {S.}~\bibnamefont {Walsh}}, \bibinfo
  {author} {\bibfnamefont {B.}~\bibnamefont {Berzin}},\ and\ \bibinfo {author}
  {\bibfnamefont {D.~A.}\ \bibnamefont {Gates}},\ }\bibfield  {title} {\bibinfo
  {title} {Overview of the helios design: A practical planar coil stellarator
  fusion power plant},\ }\href
  {https://doi.org/10.1016/j.fusengdes.2026.115999} {\bibfield  {journal}
  {\bibinfo  {journal} {Fusion Eng. Des.}\ }\textbf {\bibinfo {volume} {232}},\
  \bibinfo {pages} {115999} (\bibinfo {year} {2026})}\BibitemShut {NoStop}%
\bibitem [{\citenamefont {Torsello}\ \emph {et~al.}(2025)\citenamefont
  {Torsello}, \citenamefont {Celentano}, \citenamefont {Civale}, \citenamefont
  {Corato}, \citenamefont {Eisterer}, \citenamefont {Gambino}, \citenamefont
  {Murphy}, \citenamefont {Speller},\ and\ \citenamefont
  {Laviano}}]{torsello25}%
  \BibitemOpen
  \bibfield  {author} {\bibinfo {author} {\bibfnamefont {D.}~\bibnamefont
  {Torsello}}, \bibinfo {author} {\bibfnamefont {G.}~\bibnamefont {Celentano}},
  \bibinfo {author} {\bibfnamefont {L.}~\bibnamefont {Civale}}, \bibinfo
  {author} {\bibfnamefont {V.}~\bibnamefont {Corato}}, \bibinfo {author}
  {\bibfnamefont {M.}~\bibnamefont {Eisterer}}, \bibinfo {author}
  {\bibfnamefont {D.}~\bibnamefont {Gambino}}, \bibinfo {author} {\bibfnamefont
  {S.}~\bibnamefont {Murphy}}, \bibinfo {author} {\bibfnamefont
  {S.}~\bibnamefont {Speller}},\ and\ \bibinfo {author} {\bibfnamefont
  {F.}~\bibnamefont {Laviano}},\ }\bibfield  {title} {\bibinfo {title} {Roadmap
  for the investigation of irradiation effects in {HTS} for fusion},\ }\href
  {https://doi.org/10.1088/1361-6668/adce40} {\bibfield  {journal} {\bibinfo
  {journal} {Supercond. Sci. Technol.}\ }\textbf {\bibinfo {volume} {38}},\
  \bibinfo {pages} {053501} (\bibinfo {year} {2025})}\BibitemShut {NoStop}%
\bibitem [{rtn(1983)}]{rtns83}%
  \BibitemOpen
  \href {https://doi.org/10.2172/6911772} {\emph {\bibinfo {title} {{RTNS-II}:
  irradiations at the Rotating Target Neutron Source-II, 1983 annual
  report}}},\ \bibinfo {type} {Tech. Rep.}\ \bibinfo {number} {UCID-19837-83}\
  (\bibinfo  {institution} {Lawrence Livermore National Laboratory},\ \bibinfo
  {year} {1983})\BibitemShut {NoStop}%
\bibitem [{\citenamefont {Guinan}\ \emph {et~al.}(1984)\citenamefont {Guinan},
  \citenamefont {Van~Konynenburg},\ and\ \citenamefont {Mitchell}}]{guinan84}%
  \BibitemOpen
  \bibfield  {author} {\bibinfo {author} {\bibfnamefont {M.~W.}\ \bibnamefont
  {Guinan}}, \bibinfo {author} {\bibfnamefont {R.~A.}\ \bibnamefont
  {Van~Konynenburg}},\ and\ \bibinfo {author} {\bibfnamefont {J.~B.}\
  \bibnamefont {Mitchell}},\ }\href {https://doi.org/10.2172/5117111} {\emph
  {\bibinfo {title} {Effects of low-temperature fusion neutron irradiation on
  critical properties of a monofilament niobium-tin superconductor}}},\
  \bibinfo {type} {Tech. Rep.}\ \bibinfo {number} {UCID-20048}\ (\bibinfo
  {institution} {Lawrence Livermore National Laboratory},\ \bibinfo {year}
  {1984})\BibitemShut {NoStop}%
\bibitem [{\citenamefont {Fischer}\ \emph {et~al.}(2015)\citenamefont
  {Fischer}, \citenamefont {Bachmann}, \citenamefont {Palermo}, \citenamefont
  {Pereslavtsev},\ and\ \citenamefont {Villari}}]{fischer15}%
  \BibitemOpen
  \bibfield  {author} {\bibinfo {author} {\bibfnamefont {U.}~\bibnamefont
  {Fischer}}, \bibinfo {author} {\bibfnamefont {C.}~\bibnamefont {Bachmann}},
  \bibinfo {author} {\bibfnamefont {I.}~\bibnamefont {Palermo}}, \bibinfo
  {author} {\bibfnamefont {P.}~\bibnamefont {Pereslavtsev}},\ and\ \bibinfo
  {author} {\bibfnamefont {R.}~\bibnamefont {Villari}},\ }\bibfield  {title}
  {\bibinfo {title} {Neutronics requirements for a {DEMO} fusion power plant},\
  }\href {https://doi.org/10.1016/j.fusengdes.2015.02.029} {\bibfield
  {journal} {\bibinfo  {journal} {Fusion Eng. Des.}\ }\textbf {\bibinfo
  {volume} {98-99}},\ \bibinfo {pages} {2134} (\bibinfo {year}
  {2015})}\BibitemShut {NoStop}%
\bibitem [{\citenamefont {{\v{C}}ufar}\ \emph {et~al.}(2021)\citenamefont
  {{\v{C}}ufar}, \citenamefont {Bachmann}, \citenamefont {Berry}, \citenamefont
  {Chavan}, \citenamefont {Eade}, \citenamefont {Franke}, \citenamefont {Kos},
  \citenamefont {Leichtle}, \citenamefont {Spaeh},\ and\ \citenamefont
  {Tran}}]{cufar21}%
  \BibitemOpen
  \bibfield  {author} {\bibinfo {author} {\bibfnamefont {A.}~\bibnamefont
  {{\v{C}}ufar}}, \bibinfo {author} {\bibfnamefont {C.}~\bibnamefont
  {Bachmann}}, \bibinfo {author} {\bibfnamefont {T.}~\bibnamefont {Berry}},
  \bibinfo {author} {\bibfnamefont {R.}~\bibnamefont {Chavan}}, \bibinfo
  {author} {\bibfnamefont {T.}~\bibnamefont {Eade}}, \bibinfo {author}
  {\bibfnamefont {T.}~\bibnamefont {Franke}}, \bibinfo {author} {\bibfnamefont
  {B.}~\bibnamefont {Kos}}, \bibinfo {author} {\bibfnamefont {D.}~\bibnamefont
  {Leichtle}}, \bibinfo {author} {\bibfnamefont {P.}~\bibnamefont {Spaeh}},\
  and\ \bibinfo {author} {\bibfnamefont {M.~Q.}\ \bibnamefont {Tran}},\
  }\bibfield  {title} {\bibinfo {title} {Neutronics analyses of {EU DEMO} 2020
  {EC} port configuration},\ }in\ \href
  {https://www.djs.si/nene2021/proceedings/pdf/NENE2021_804.pdf} {\emph
  {\bibinfo {booktitle} {Proceedings of the 30th International Conference
  Nuclear Energy for New Europe ({NENE} 2021)}}}\ (\bibinfo {address} {Bled,
  Slovenia},\ \bibinfo {year} {2021})\ p.\ \bibinfo {pages} {804}\BibitemShut
  {NoStop}%
\bibitem [{\citenamefont {Senatore}\ \emph {et~al.}(2024)\citenamefont
  {Senatore}, \citenamefont {Bonura},\ and\ \citenamefont
  {Bagni}}]{senatore24}%
  \BibitemOpen
  \bibfield  {author} {\bibinfo {author} {\bibfnamefont {C.}~\bibnamefont
  {Senatore}}, \bibinfo {author} {\bibfnamefont {M.}~\bibnamefont {Bonura}},\
  and\ \bibinfo {author} {\bibfnamefont {T.}~\bibnamefont {Bagni}},\ }\bibfield
   {title} {\bibinfo {title} {{REBCO} tapes for applications in ultra-high
  fields: critical current surface and scaling relations},\ }\href
  {https://doi.org/10.1088/1361-6668/ad7f95} {\bibfield  {journal} {\bibinfo
  {journal} {Supercond. Sci. Technol.}\ }\textbf {\bibinfo {volume} {37}},\
  \bibinfo {pages} {115013} (\bibinfo {year} {2024})}\BibitemShut {NoStop}%
\bibitem [{\citenamefont {Chislett-McDonald}\ \emph {et~al.}(2022)\citenamefont
  {Chislett-McDonald}, \citenamefont {Surrey}, \citenamefont {Naish},
  \citenamefont {Turner},\ and\ \citenamefont {Hampshire}}]{chislett22}%
  \BibitemOpen
  \bibfield  {author} {\bibinfo {author} {\bibfnamefont {S.~B.~L.}\
  \bibnamefont {Chislett-McDonald}}, \bibinfo {author} {\bibfnamefont
  {E.}~\bibnamefont {Surrey}}, \bibinfo {author} {\bibfnamefont
  {J.}~\bibnamefont {Naish}}, \bibinfo {author} {\bibfnamefont
  {A.}~\bibnamefont {Turner}},\ and\ \bibinfo {author} {\bibfnamefont {D.~P.}\
  \bibnamefont {Hampshire}},\ }\href@noop {} {\bibinfo {title} {Training and
  upgrading tokamak power plants with remountable superconducting magnets}}
  (\bibinfo {year} {2022}),\ \Eprint {https://arxiv.org/abs/2205.04441}
  {arXiv:2205.04441 [physics.plasm-ph]} \BibitemShut {NoStop}%
\bibitem [{\citenamefont {Lyon}\ \emph {et~al.}(1994)\citenamefont {Lyon},
  \citenamefont {Gulec}, \citenamefont {Miller},\ and\ \citenamefont
  {El-Guebaly}}]{lyon94}%
  \BibitemOpen
  \bibfield  {author} {\bibinfo {author} {\bibfnamefont {J.~F.}\ \bibnamefont
  {Lyon}}, \bibinfo {author} {\bibfnamefont {K.}~\bibnamefont {Gulec}},
  \bibinfo {author} {\bibfnamefont {R.~L.}\ \bibnamefont {Miller}},\ and\
  \bibinfo {author} {\bibfnamefont {L.}~\bibnamefont {El-Guebaly}},\ }\href
  {https://doi.org/10.2172/10141687} {\emph {\bibinfo {title} {Evaluation of
  torsatrons as reactors}}},\ \bibinfo {type} {Tech. Rep.}\ \bibinfo {number}
  {ORNL/TM-12670}\ (\bibinfo  {institution} {Oak Ridge National Laboratory},\
  \bibinfo {year} {1994})\BibitemShut {NoStop}%
\bibitem [{\citenamefont {Lyon}\ \emph {et~al.}(1997)\citenamefont {Lyon},
  \citenamefont {Anderson}, \citenamefont {Rome}, \citenamefont {Johnson},\
  and\ \citenamefont {Garabedian}}]{spps97}%
  \BibitemOpen
  \bibfield  {author} {\bibinfo {author} {\bibfnamefont {J.~F.}\ \bibnamefont
  {Lyon}}, \bibinfo {author} {\bibfnamefont {D.~T.}\ \bibnamefont {Anderson}},
  \bibinfo {author} {\bibfnamefont {J.~A.}\ \bibnamefont {Rome}}, \bibinfo
  {author} {\bibfnamefont {J.~L.}\ \bibnamefont {Johnson}},\ and\ \bibinfo
  {author} {\bibfnamefont {P.~R.}\ \bibnamefont {Garabedian}},\ }\href
  {https://qedfusion.org/LIB/REPORT/SPPS/FINAL/chap2.pdf} {\emph {\bibinfo
  {title} {Stellarator physics}}},\ \bibinfo {type} {Tech. Rep.}\ \bibinfo
  {number} {UCSD-ENG-004}\ (\bibinfo  {institution} {University of California,
  San Diego},\ \bibinfo {year} {1997})\ \bibinfo {note} {chapter 2 of the
  Stellarator Power Plant Study final report}\BibitemShut {NoStop}%
\bibitem [{\citenamefont {Ku}\ \emph {et~al.}(2008)\citenamefont {Ku},
  \citenamefont {Garabedian}, \citenamefont {Lyon}, \citenamefont {Turnbull},
  \citenamefont {Grossman}, \citenamefont {Mau}, \citenamefont {Zarnstorff},\
  and\ \citenamefont {{ARIES Team}}}]{ku08}%
  \BibitemOpen
  \bibfield  {author} {\bibinfo {author} {\bibfnamefont {L.~P.}\ \bibnamefont
  {Ku}}, \bibinfo {author} {\bibfnamefont {P.~R.}\ \bibnamefont {Garabedian}},
  \bibinfo {author} {\bibfnamefont {J.}~\bibnamefont {Lyon}}, \bibinfo {author}
  {\bibfnamefont {A.}~\bibnamefont {Turnbull}}, \bibinfo {author}
  {\bibfnamefont {A.}~\bibnamefont {Grossman}}, \bibinfo {author}
  {\bibfnamefont {T.~K.}\ \bibnamefont {Mau}}, \bibinfo {author} {\bibfnamefont
  {M.}~\bibnamefont {Zarnstorff}},\ and\ \bibinfo {author} {\bibnamefont
  {{ARIES Team}}},\ }\bibfield  {title} {\bibinfo {title} {Physics design for
  {ARIES-CS}},\ }\href {https://doi.org/10.13182/FST08-A1899} {\bibfield
  {journal} {\bibinfo  {journal} {Fusion Sci. Technol.}\ }\textbf {\bibinfo
  {volume} {54}},\ \bibinfo {pages} {673} (\bibinfo {year} {2008})}\BibitemShut
  {NoStop}%
\bibitem [{\citenamefont {Kappel}\ \emph {et~al.}(2024)\citenamefont {Kappel},
  \citenamefont {Landreman},\ and\ \citenamefont {Malhotra}}]{kappel24}%
  \BibitemOpen
  \bibfield  {author} {\bibinfo {author} {\bibfnamefont {J.}~\bibnamefont
  {Kappel}}, \bibinfo {author} {\bibfnamefont {M.}~\bibnamefont {Landreman}},\
  and\ \bibinfo {author} {\bibfnamefont {D.}~\bibnamefont {Malhotra}},\
  }\bibfield  {title} {\bibinfo {title} {The magnetic gradient scale length
  explains why certain plasmas require close external magnetic coils},\ }\href
  {https://doi.org/10.1088/1361-6587/ad1a3e} {\bibfield  {journal} {\bibinfo
  {journal} {Plasma Phys. Control. Fusion}\ }\textbf {\bibinfo {volume} {66}},\
  \bibinfo {pages} {025018} (\bibinfo {year} {2024})}\BibitemShut {NoStop}%
\bibitem [{\citenamefont {Lyytinen}\ \emph {et~al.}(2024)\citenamefont
  {Lyytinen}, \citenamefont {Snicker}, \citenamefont {Virtanen}, \citenamefont
  {Palermo}, \citenamefont {Alguacil}, \citenamefont {Bogaarts},\ and\
  \citenamefont {Warmer}}]{lyytinen24}%
  \BibitemOpen
  \bibfield  {author} {\bibinfo {author} {\bibfnamefont {T.}~\bibnamefont
  {Lyytinen}}, \bibinfo {author} {\bibfnamefont {A.}~\bibnamefont {Snicker}},
  \bibinfo {author} {\bibfnamefont {J.}~\bibnamefont {Virtanen}}, \bibinfo
  {author} {\bibfnamefont {I.}~\bibnamefont {Palermo}}, \bibinfo {author}
  {\bibfnamefont {J.}~\bibnamefont {Alguacil}}, \bibinfo {author}
  {\bibfnamefont {T.}~\bibnamefont {Bogaarts}},\ and\ \bibinfo {author}
  {\bibfnamefont {F.}~\bibnamefont {Warmer}},\ }\bibfield  {title} {\bibinfo
  {title} {Proof-of-principle of parametric stellarator neutronics modeling
  using {Serpent2}},\ }\href {https://doi.org/10.1088/1741-4326/ad4f9f}
  {\bibfield  {journal} {\bibinfo  {journal} {Nucl. Fusion}\ }\textbf {\bibinfo
  {volume} {64}},\ \bibinfo {pages} {076042} (\bibinfo {year}
  {2024})}\BibitemShut {NoStop}%
\bibitem [{\citenamefont {Wiedman}\ \emph {et~al.}(2024)\citenamefont
  {Wiedman}, \citenamefont {Buller},\ and\ \citenamefont
  {Landreman}}]{Wiedman2024}%
  \BibitemOpen
  \bibfield  {author} {\bibinfo {author} {\bibfnamefont {A.}~\bibnamefont
  {Wiedman}}, \bibinfo {author} {\bibfnamefont {S.}~\bibnamefont {Buller}},\
  and\ \bibinfo {author} {\bibfnamefont {M.}~\bibnamefont {Landreman}},\
  }\bibfield  {title} {\bibinfo {title} {Coil optimization for quasi-helically
  symmetric stellarator configurations},\ }\href
  {https://doi.org/10.1017/S0022377824000540} {\bibfield  {journal} {\bibinfo
  {journal} {J. Plasma Phys.}\ }\textbf {\bibinfo {volume} {90}},\ \bibinfo
  {pages} {905900307} (\bibinfo {year} {2024})}\BibitemShut {NoStop}%
\bibitem [{\citenamefont {Gil}\ \emph {et~al.}(2026)\citenamefont {Gil},
  \citenamefont {Li}, \citenamefont {Stratton}, \citenamefont {Kaptanoglu},\
  and\ \citenamefont {Stenson}}]{gil25}%
  \BibitemOpen
  \bibfield  {author} {\bibinfo {author} {\bibfnamefont {P.~F.}\ \bibnamefont
  {Gil}}, \bibinfo {author} {\bibfnamefont {W.}~\bibnamefont {Li}}, \bibinfo
  {author} {\bibfnamefont {J.}~\bibnamefont {Stratton}}, \bibinfo {author}
  {\bibfnamefont {A.~A.}\ \bibnamefont {Kaptanoglu}},\ and\ \bibinfo {author}
  {\bibfnamefont {E.~V.}\ \bibnamefont {Stenson}},\ }\bibfield  {title}
  {\bibinfo {title} {Efficient computation of stellarator coils with an
  augmented {L}agrangian optimization method},\ }\href
  {https://doi.org/10.1103/838f-vbq3} {\bibfield  {journal} {\bibinfo
  {journal} {Phys. Rev. E}\ }\textbf {\bibinfo {volume} {114}},\ \bibinfo
  {pages} {025202} (\bibinfo {year} {2026})}\BibitemShut {NoStop}%
\bibitem [{\citenamefont {Ewers}\ \emph {et~al.}(2026)\citenamefont {Ewers},
  \citenamefont {Cadena}, \citenamefont {Foppiani},\ and\ \citenamefont
  {Merlo}}]{ewers26}%
  \BibitemOpen
  \bibfield  {author} {\bibinfo {author} {\bibfnamefont {J.-H.}\ \bibnamefont
  {Ewers}}, \bibinfo {author} {\bibfnamefont {S.~A.}\ \bibnamefont {Cadena}},
  \bibinfo {author} {\bibfnamefont {N.}~\bibnamefont {Foppiani}},\ and\
  \bibinfo {author} {\bibfnamefont {A.}~\bibnamefont {Merlo}},\ }\href@noop {}
  {\bibinfo {title} {{CoilStellaration}: A dataset and benchmark for
  engineering-aware stellarator coilset generation}} (\bibinfo {year} {2026}),\
  \Eprint {https://arxiv.org/abs/2611.12689} {arXiv:2611.12689} \BibitemShut
  {NoStop}%
\bibitem [{\citenamefont {Kappel}\ \emph {et~al.}(2026)\citenamefont {Kappel},
  \citenamefont {Landreman}, \citenamefont {Jura{\v{s}}i{\'c}},\ and\
  \citenamefont {Henneberg}}]{kappel26}%
  \BibitemOpen
  \bibfield  {author} {\bibinfo {author} {\bibfnamefont {J.}~\bibnamefont
  {Kappel}}, \bibinfo {author} {\bibfnamefont {M.}~\bibnamefont {Landreman}},
  \bibinfo {author} {\bibfnamefont {P.}~\bibnamefont {Jura{\v{s}}i{\'c}}},\
  and\ \bibinfo {author} {\bibfnamefont {S.~A.}\ \bibnamefont {Henneberg}},\
  }\bibfield  {title} {\bibinfo {title} {How does the magnetic gradient scale
  length influence complexity of filamentary coils in stellarators?},\ }\href
  {https://doi.org/10.1088/1741-4326/ae69fa} {\bibfield  {journal} {\bibinfo
  {journal} {Nucl. Fusion}\ }\textbf {\bibinfo {volume} {66}},\ \bibinfo
  {pages} {066021} (\bibinfo {year} {2026})}\BibitemShut {NoStop}%
\bibitem [{\citenamefont {Henneberg}\ and\ \citenamefont
  {Zettl}(2026)}]{henneberg26}%
  \BibitemOpen
  \bibfield  {author} {\bibinfo {author} {\bibfnamefont {S.~A.}\ \bibnamefont
  {Henneberg}}\ and\ \bibinfo {author} {\bibfnamefont {A.}~\bibnamefont
  {Zettl}},\ }\href@noop {} {\bibinfo {title} {A new degree of freedom in
  stellarator coil design}} (\bibinfo {year} {2026}),\ \Eprint
  {https://arxiv.org/abs/2610.14396} {arXiv:2610.14396} \BibitemShut {NoStop}%
\bibitem [{\citenamefont {Paul}\ \emph {et~al.}(2018)\citenamefont {Paul},
  \citenamefont {Landreman}, \citenamefont {Bader},\ and\ \citenamefont
  {Dorland}}]{paul18}%
  \BibitemOpen
  \bibfield  {author} {\bibinfo {author} {\bibfnamefont {E.~J.}\ \bibnamefont
  {Paul}}, \bibinfo {author} {\bibfnamefont {M.}~\bibnamefont {Landreman}},
  \bibinfo {author} {\bibfnamefont {A.}~\bibnamefont {Bader}},\ and\ \bibinfo
  {author} {\bibfnamefont {W.}~\bibnamefont {Dorland}},\ }\bibfield  {title}
  {\bibinfo {title} {An adjoint method for gradient-based optimization of
  stellarator coil shapes},\ }\href {https://doi.org/10.1088/1741-4326/aac1c7}
  {\bibfield  {journal} {\bibinfo  {journal} {Nucl. Fusion}\ }\textbf {\bibinfo
  {volume} {58}},\ \bibinfo {pages} {076015} (\bibinfo {year}
  {2018})}\BibitemShut {NoStop}%
\bibitem [{\citenamefont {Strickler}\ \emph {et~al.}(2002)\citenamefont
  {Strickler}, \citenamefont {Berry}, \citenamefont {Hirshman} \emph
  {et~al.}}]{strickler02}%
  \BibitemOpen
  \bibfield  {author} {\bibinfo {author} {\bibfnamefont {D.~J.}\ \bibnamefont
  {Strickler}}, \bibinfo {author} {\bibfnamefont {L.~A.}\ \bibnamefont
  {Berry}}, \bibinfo {author} {\bibfnamefont {S.~P.}\ \bibnamefont {Hirshman}},
  \emph {et~al.},\ }\bibfield  {title} {\bibinfo {title} {Integrated plasma and
  coil optimization for compact stellarators},\ }in\ \href
  {https://www-pub.iaea.org/MTCD/publications/PDF/csp_019c/pdf/ftp2_06.pdf}
  {\emph {\bibinfo {booktitle} {19th IAEA Fusion Energy Conference, Lyon}}}\
  (\bibinfo {year} {2002})\ \bibinfo {note} {{FT/P2-06}}\BibitemShut {NoStop}%
\bibitem [{\citenamefont {Henneberg}\ \emph {et~al.}(2021)\citenamefont
  {Henneberg}, \citenamefont {Hudson}, \citenamefont {Pfefferl{\'e}},\ and\
  \citenamefont {Helander}}]{henneberg21}%
  \BibitemOpen
  \bibfield  {author} {\bibinfo {author} {\bibfnamefont {S.~A.}\ \bibnamefont
  {Henneberg}}, \bibinfo {author} {\bibfnamefont {S.~R.}\ \bibnamefont
  {Hudson}}, \bibinfo {author} {\bibfnamefont {D.}~\bibnamefont
  {Pfefferl{\'e}}},\ and\ \bibinfo {author} {\bibfnamefont {P.}~\bibnamefont
  {Helander}},\ }\bibfield  {title} {\bibinfo {title} {Combined plasma-coil
  optimization algorithms},\ }\href {https://doi.org/10.1017/s0022377821000271}
  {\bibfield  {journal} {\bibinfo  {journal} {J. Plasma Phys.}\ }\textbf
  {\bibinfo {volume} {87}},\ \bibinfo {pages} {905870226} (\bibinfo {year}
  {2021})}\BibitemShut {NoStop}%
\bibitem [{\citenamefont {Giuliani}\ \emph {et~al.}(2022)\citenamefont
  {Giuliani}, \citenamefont {Wechsung}, \citenamefont {Cerfon}, \citenamefont
  {Stadler},\ and\ \citenamefont {Landreman}}]{giuliani22}%
  \BibitemOpen
  \bibfield  {author} {\bibinfo {author} {\bibfnamefont {A.}~\bibnamefont
  {Giuliani}}, \bibinfo {author} {\bibfnamefont {F.}~\bibnamefont {Wechsung}},
  \bibinfo {author} {\bibfnamefont {A.}~\bibnamefont {Cerfon}}, \bibinfo
  {author} {\bibfnamefont {G.}~\bibnamefont {Stadler}},\ and\ \bibinfo {author}
  {\bibfnamefont {M.}~\bibnamefont {Landreman}},\ }\bibfield  {title} {\bibinfo
  {title} {Single-stage gradient-based stellarator coil design: Optimization
  for near-axis quasi-symmetry},\ }\href
  {https://doi.org/10.1016/j.jcp.2022.111147} {\bibfield  {journal} {\bibinfo
  {journal} {J. Comput. Phys.}\ }\textbf {\bibinfo {volume} {459}},\ \bibinfo
  {pages} {111147} (\bibinfo {year} {2022})}\BibitemShut {NoStop}%
\bibitem [{\citenamefont {Jorge}\ \emph {et~al.}(2023)\citenamefont {Jorge},
  \citenamefont {Goodman}, \citenamefont {Landreman}, \citenamefont
  {Rodrigues},\ and\ \citenamefont {Wechsung}}]{jorge23}%
  \BibitemOpen
  \bibfield  {author} {\bibinfo {author} {\bibfnamefont {R.}~\bibnamefont
  {Jorge}}, \bibinfo {author} {\bibfnamefont {A.}~\bibnamefont {Goodman}},
  \bibinfo {author} {\bibfnamefont {M.}~\bibnamefont {Landreman}}, \bibinfo
  {author} {\bibfnamefont {J.}~\bibnamefont {Rodrigues}},\ and\ \bibinfo
  {author} {\bibfnamefont {F.}~\bibnamefont {Wechsung}},\ }\bibfield  {title}
  {\bibinfo {title} {Single-stage stellarator optimization: combining coils
  with fixed boundary equilibria},\ }\href
  {https://doi.org/10.1088/1361-6587/acd957} {\bibfield  {journal} {\bibinfo
  {journal} {Plasma Phys. Control. Fusion}\ }\textbf {\bibinfo {volume} {65}},\
  \bibinfo {pages} {074003} (\bibinfo {year} {2023})}\BibitemShut {NoStop}%
\bibitem [{\citenamefont {Fu}\ \emph {et~al.}(2026)\citenamefont {Fu},
  \citenamefont {Panici}, \citenamefont {Paul}, \citenamefont {Kaptanoglu},\
  and\ \citenamefont {Bhattacharjee}}]{fu26}%
  \BibitemOpen
  \bibfield  {author} {\bibinfo {author} {\bibfnamefont {L.}~\bibnamefont
  {Fu}}, \bibinfo {author} {\bibfnamefont {D.}~\bibnamefont {Panici}}, \bibinfo
  {author} {\bibfnamefont {E.~J.}\ \bibnamefont {Paul}}, \bibinfo {author}
  {\bibfnamefont {A.~A.}\ \bibnamefont {Kaptanoglu}},\ and\ \bibinfo {author}
  {\bibfnamefont {A.}~\bibnamefont {Bhattacharjee}},\ }\bibfield  {title}
  {\bibinfo {title} {A flexible and differentiable coil proxy for stellarator
  equilibrium optimization},\ }\href {https://doi.org/10.1063/5.0307710}
  {\bibfield  {journal} {\bibinfo  {journal} {Phys. Plasmas}\ }\textbf
  {\bibinfo {volume} {33}},\ \bibinfo {pages} {062506} (\bibinfo {year}
  {2026})}\BibitemShut {NoStop}%
\bibitem [{\citenamefont {Yu}\ \emph {et~al.}(2026)\citenamefont {Yu},
  \citenamefont {Xie}, \citenamefont {Liu},\ and\ \citenamefont {Zhu}}]{yu26}%
  \BibitemOpen
  \bibfield  {author} {\bibinfo {author} {\bibfnamefont {G.}~\bibnamefont
  {Yu}}, \bibinfo {author} {\bibfnamefont {Y.}~\bibnamefont {Xie}}, \bibinfo
  {author} {\bibfnamefont {H.}~\bibnamefont {Liu}},\ and\ \bibinfo {author}
  {\bibfnamefont {C.}~\bibnamefont {Zhu}},\ }\href@noop {} {\bibinfo {title}
  {Quasi-single-stage optimization for advanced stellarators}} (\bibinfo {year}
  {2026}),\ \Eprint {https://arxiv.org/abs/2608.03122} {arXiv:2608.03122}
  \BibitemShut {NoStop}%
\bibitem [{\citenamefont {Pomphrey}\ \emph {et~al.}(2001)\citenamefont
  {Pomphrey}, \citenamefont {Berry}, \citenamefont {Boozer}, \citenamefont
  {Brooks}, \citenamefont {Hatcher}, \citenamefont {Hirshman}, \citenamefont
  {Ku}, \citenamefont {Miner}, \citenamefont {Mynick}, \citenamefont
  {Reiersen}, \citenamefont {Strickler},\ and\ \citenamefont
  {Valanju}}]{pomphrey01}%
  \BibitemOpen
  \bibfield  {author} {\bibinfo {author} {\bibfnamefont {N.}~\bibnamefont
  {Pomphrey}}, \bibinfo {author} {\bibfnamefont {L.}~\bibnamefont {Berry}},
  \bibinfo {author} {\bibfnamefont {A.}~\bibnamefont {Boozer}}, \bibinfo
  {author} {\bibfnamefont {A.}~\bibnamefont {Brooks}}, \bibinfo {author}
  {\bibfnamefont {R.~E.}\ \bibnamefont {Hatcher}}, \bibinfo {author}
  {\bibfnamefont {S.~P.}\ \bibnamefont {Hirshman}}, \bibinfo {author}
  {\bibfnamefont {L.-P.}\ \bibnamefont {Ku}}, \bibinfo {author} {\bibfnamefont
  {W.~H.}\ \bibnamefont {Miner}}, \bibinfo {author} {\bibfnamefont {H.~E.}\
  \bibnamefont {Mynick}}, \bibinfo {author} {\bibfnamefont {W.}~\bibnamefont
  {Reiersen}}, \bibinfo {author} {\bibfnamefont {D.~J.}\ \bibnamefont
  {Strickler}},\ and\ \bibinfo {author} {\bibfnamefont {P.~M.}\ \bibnamefont
  {Valanju}},\ }\bibfield  {title} {\bibinfo {title} {Innovations in compact
  stellarator coil design},\ }\href
  {https://doi.org/10.1088/0029-5515/41/3/312} {\bibfield  {journal} {\bibinfo
  {journal} {Nucl. Fusion}\ }\textbf {\bibinfo {volume} {41}},\ \bibinfo
  {pages} {339} (\bibinfo {year} {2001})}\BibitemShut {NoStop}%
\bibitem [{\citenamefont {Carlton-Jones}\ \emph {et~al.}(2021)\citenamefont
  {Carlton-Jones}, \citenamefont {Paul},\ and\ \citenamefont
  {Dorland}}]{carltonjones21}%
  \BibitemOpen
  \bibfield  {author} {\bibinfo {author} {\bibfnamefont {A.}~\bibnamefont
  {Carlton-Jones}}, \bibinfo {author} {\bibfnamefont {E.~J.}\ \bibnamefont
  {Paul}},\ and\ \bibinfo {author} {\bibfnamefont {W.}~\bibnamefont
  {Dorland}},\ }\bibfield  {title} {\bibinfo {title} {Computing the shape
  gradient of stellarator coil complexity with respect to the plasma
  boundary},\ }\href {https://doi.org/10.1017/S0022377821000386} {\bibfield
  {journal} {\bibinfo  {journal} {J. Plasma Phys.}\ }\textbf {\bibinfo {volume}
  {87}},\ \bibinfo {pages} {905870222} (\bibinfo {year} {2021})}\BibitemShut
  {NoStop}%
\bibitem [{\citenamefont {Landreman}\ and\ \citenamefont
  {Boozer}(2016)}]{landreman16}%
  \BibitemOpen
  \bibfield  {author} {\bibinfo {author} {\bibfnamefont {M.}~\bibnamefont
  {Landreman}}\ and\ \bibinfo {author} {\bibfnamefont {A.~H.}\ \bibnamefont
  {Boozer}},\ }\bibfield  {title} {\bibinfo {title} {Efficient magnetic fields
  for supporting toroidal plasmas},\ }\href {https://doi.org/10.1063/1.4943201}
  {\bibfield  {journal} {\bibinfo  {journal} {Phys. Plasmas}\ }\textbf
  {\bibinfo {volume} {23}},\ \bibinfo {pages} {032506} (\bibinfo {year}
  {2016})}\BibitemShut {NoStop}%
\bibitem [{\citenamefont {Goodman}\ \emph {et~al.}(2024)\citenamefont
  {Goodman}, \citenamefont {Xanthopoulos}, \citenamefont {Plunk}, \citenamefont
  {Smith}, \citenamefont {N{\"u}hrenberg}, \citenamefont {Beidler},
  \citenamefont {Henneberg}, \citenamefont {Roberg-Clark}, \citenamefont
  {Drevlak},\ and\ \citenamefont {Helander}}]{Goodman2024}%
  \BibitemOpen
  \bibfield  {author} {\bibinfo {author} {\bibfnamefont {A.~G.}\ \bibnamefont
  {Goodman}}, \bibinfo {author} {\bibfnamefont {P.}~\bibnamefont
  {Xanthopoulos}}, \bibinfo {author} {\bibfnamefont {G.~G.}\ \bibnamefont
  {Plunk}}, \bibinfo {author} {\bibfnamefont {H.}~\bibnamefont {Smith}},
  \bibinfo {author} {\bibfnamefont {C.}~\bibnamefont {N{\"u}hrenberg}},
  \bibinfo {author} {\bibfnamefont {C.~D.}\ \bibnamefont {Beidler}}, \bibinfo
  {author} {\bibfnamefont {S.~A.}\ \bibnamefont {Henneberg}}, \bibinfo {author}
  {\bibfnamefont {G.}~\bibnamefont {Roberg-Clark}}, \bibinfo {author}
  {\bibfnamefont {M.}~\bibnamefont {Drevlak}},\ and\ \bibinfo {author}
  {\bibfnamefont {P.}~\bibnamefont {Helander}},\ }\bibfield  {title} {\bibinfo
  {title} {Quasi-isodynamic stellarators with low turbulence as fusion reactor
  candidates},\ }\href {https://doi.org/10.1103/prxenergy.3.023010} {\bibfield
  {journal} {\bibinfo  {journal} {PRX Energy}\ }\textbf {\bibinfo {volume}
  {3}},\ \bibinfo {pages} {023010} (\bibinfo {year} {2024})}\BibitemShut
  {NoStop}%
\bibitem [{\citenamefont {Goto}\ \emph {et~al.}(2011)\citenamefont {Goto},
  \citenamefont {Suzuki}, \citenamefont {Yanagi}, \citenamefont {Watanabe},
  \citenamefont {Imagawa},\ and\ \citenamefont {Sagara}}]{goto11}%
  \BibitemOpen
  \bibfield  {author} {\bibinfo {author} {\bibfnamefont {T.}~\bibnamefont
  {Goto}}, \bibinfo {author} {\bibfnamefont {Y.}~\bibnamefont {Suzuki}},
  \bibinfo {author} {\bibfnamefont {N.}~\bibnamefont {Yanagi}}, \bibinfo
  {author} {\bibfnamefont {K.~Y.}\ \bibnamefont {Watanabe}}, \bibinfo {author}
  {\bibfnamefont {S.}~\bibnamefont {Imagawa}},\ and\ \bibinfo {author}
  {\bibfnamefont {A.}~\bibnamefont {Sagara}},\ }\bibfield  {title} {\bibinfo
  {title} {Importance of helical pitch parameter in {LHD}-type heliotron
  reactor designs},\ }\href {https://doi.org/10.1088/0029-5515/51/8/083045}
  {\bibfield  {journal} {\bibinfo  {journal} {Nucl. Fusion}\ }\textbf {\bibinfo
  {volume} {51}},\ \bibinfo {pages} {083045} (\bibinfo {year}
  {2011})}\BibitemShut {NoStop}%
\bibitem [{\citenamefont {Yanagi}\ \emph {et~al.}(2016)\citenamefont {Yanagi},
  \citenamefont {Goto}, \citenamefont {Tamura}, \citenamefont {Miyazawa},\ and\
  \citenamefont {Sagara}}]{yanagi16}%
  \BibitemOpen
  \bibfield  {author} {\bibinfo {author} {\bibfnamefont {N.}~\bibnamefont
  {Yanagi}}, \bibinfo {author} {\bibfnamefont {T.}~\bibnamefont {Goto}},
  \bibinfo {author} {\bibfnamefont {H.}~\bibnamefont {Tamura}}, \bibinfo
  {author} {\bibfnamefont {J.}~\bibnamefont {Miyazawa}},\ and\ \bibinfo
  {author} {\bibfnamefont {A.}~\bibnamefont {Sagara}},\ }\bibfield  {title}
  {\bibinfo {title} {{NITA} coil: innovation for enlarging the blanket space in
  the helical fusion reactor},\ }\href {https://doi.org/10.1585/pfr.11.2405034}
  {\bibfield  {journal} {\bibinfo  {journal} {Plasma Fusion Res.}\ }\textbf
  {\bibinfo {volume} {11}},\ \bibinfo {pages} {2405034} (\bibinfo {year}
  {2016})}\BibitemShut {NoStop}%
\bibitem [{\citenamefont {Moreno}\ \emph {et~al.}(2024)\citenamefont {Moreno},
  \citenamefont {Bader},\ and\ \citenamefont {Wilson}}]{moreno24}%
  \BibitemOpen
  \bibfield  {author} {\bibinfo {author} {\bibfnamefont {C.}~\bibnamefont
  {Moreno}}, \bibinfo {author} {\bibfnamefont {A.}~\bibnamefont {Bader}},\ and\
  \bibinfo {author} {\bibfnamefont {P.}~\bibnamefont {Wilson}},\ }\bibfield
  {title} {\bibinfo {title} {{ParaStell}: parametric modeling and neutronics
  support for stellarator fusion power plants},\ }\href
  {https://doi.org/10.3389/fnuen.2024.1384788} {\bibfield  {journal} {\bibinfo
  {journal} {Front. Nucl. Eng.}\ }\textbf {\bibinfo {volume} {3}},\ \bibinfo
  {pages} {1384788} (\bibinfo {year} {2024})}\BibitemShut {NoStop}%
\bibitem [{\citenamefont {Bogaarts}\ and\ \citenamefont
  {Warmer}(2026)}]{bogaarts26}%
  \BibitemOpen
  \bibfield  {author} {\bibinfo {author} {\bibfnamefont {T.~J.}\ \bibnamefont
  {Bogaarts}}\ and\ \bibinfo {author} {\bibfnamefont {F.}~\bibnamefont
  {Warmer}},\ }\bibfield  {title} {\bibinfo {title} {A novel
  discontinuous-{G}alerkin deterministic neutronics model for fusion
  applications: workflow for stellarator reactor design studies},\ }\href
  {https://doi.org/10.1088/1741-4326/ae4e48} {\bibfield  {journal} {\bibinfo
  {journal} {Nucl. Fusion}\ }\textbf {\bibinfo {volume} {66}},\ \bibinfo
  {pages} {046020} (\bibinfo {year} {2026})}\BibitemShut {NoStop}%
\bibitem [{\citenamefont {Ku}\ and\ \citenamefont
  {Boozer}(2010{\natexlab{a}})}]{ku10}%
  \BibitemOpen
  \bibfield  {author} {\bibinfo {author} {\bibfnamefont {L.~P.}\ \bibnamefont
  {Ku}}\ and\ \bibinfo {author} {\bibfnamefont {A.~H.}\ \bibnamefont
  {Boozer}},\ }\bibfield  {title} {\bibinfo {title} {Stellarator coil design
  and plasma sensitivity},\ }\href {https://doi.org/10.1063/1.3527994}
  {\bibfield  {journal} {\bibinfo  {journal} {Phys. Plasmas}\ }\textbf
  {\bibinfo {volume} {17}},\ \bibinfo {pages} {122503} (\bibinfo {year}
  {2010}{\natexlab{a}})}\BibitemShut {NoStop}%
\bibitem [{\citenamefont {Snead}\ \emph {et~al.}(1981)\citenamefont {Snead},
  \citenamefont {Parkin},\ and\ \citenamefont {Guinan}}]{snead81}%
  \BibitemOpen
  \bibfield  {author} {\bibinfo {author} {\bibfnamefont {C.~L.}\ \bibnamefont
  {Snead}}, \bibinfo {author} {\bibfnamefont {D.~M.}\ \bibnamefont {Parkin}},\
  and\ \bibinfo {author} {\bibfnamefont {M.~W.}\ \bibnamefont {Guinan}},\
  }\bibfield  {title} {\bibinfo {title} {High-energy-neutron damage in
  {Nb$_3$Sn}: changes in critical properties, and damage-energy analysis},\
  }\href {https://doi.org/10.1016/0022-3115(82)90688-2} {\bibfield  {journal}
  {\bibinfo  {journal} {J. Nucl. Mater.}\ }\textbf {\bibinfo {volume} {103}},\
  \bibinfo {pages} {749} (\bibinfo {year} {1981})}\BibitemShut {NoStop}%
\bibitem [{\citenamefont {Nishimura}\ \emph {et~al.}(2010)\citenamefont
  {Nishimura}, \citenamefont {Takeuchi}, \citenamefont {Nishijima},
  \citenamefont {Koizumi}, \citenamefont {Nishijima},\ and\ \citenamefont
  {Watanabe}}]{nishimura10}%
  \BibitemOpen
  \bibfield  {author} {\bibinfo {author} {\bibfnamefont {A.}~\bibnamefont
  {Nishimura}}, \bibinfo {author} {\bibfnamefont {T.}~\bibnamefont {Takeuchi}},
  \bibinfo {author} {\bibfnamefont {S.}~\bibnamefont {Nishijima}}, \bibinfo
  {author} {\bibfnamefont {N.}~\bibnamefont {Koizumi}}, \bibinfo {author}
  {\bibfnamefont {G.}~\bibnamefont {Nishijima}},\ and\ \bibinfo {author}
  {\bibfnamefont {K.}~\bibnamefont {Watanabe}},\ }\bibfield  {title} {\bibinfo
  {title} {14 {MeV} neutron irradiation effect on superconducting properties of
  {Nb$_3$Sn} strand for fusion magnet},\ }in\ \href
  {https://www-pub.iaea.org/MTCD/meetings/PDFplus/2010/cn180/cn180_papers/ftp_p1-22.pdf}
  {\emph {\bibinfo {booktitle} {23rd IAEA Fusion Energy Conference}}}\
  (\bibinfo {address} {Daejeon},\ \bibinfo {year} {2010})\ \bibinfo {note}
  {{FTP/P1-22}}\BibitemShut {NoStop}%
\bibitem [{\citenamefont {Nishimura}\ \emph {et~al.}(2008)\citenamefont
  {Nishimura}, \citenamefont {Takeuchi}, \citenamefont {Nishijima},
  \citenamefont {Izumi}, \citenamefont {Takakura}, \citenamefont {Ochiai},
  \citenamefont {Henmi}, \citenamefont {Nishijima}, \citenamefont {Watanabe},
  \citenamefont {Sato}, \citenamefont {Kurisita}, \citenamefont {Narui},\ and\
  \citenamefont {Shikama}}]{nishimura08}%
  \BibitemOpen
  \bibfield  {author} {\bibinfo {author} {\bibfnamefont {A.}~\bibnamefont
  {Nishimura}}, \bibinfo {author} {\bibfnamefont {T.}~\bibnamefont {Takeuchi}},
  \bibinfo {author} {\bibfnamefont {S.}~\bibnamefont {Nishijima}}, \bibinfo
  {author} {\bibfnamefont {Y.}~\bibnamefont {Izumi}}, \bibinfo {author}
  {\bibfnamefont {K.}~\bibnamefont {Takakura}}, \bibinfo {author}
  {\bibfnamefont {K.}~\bibnamefont {Ochiai}}, \bibinfo {author} {\bibfnamefont
  {T.}~\bibnamefont {Henmi}}, \bibinfo {author} {\bibfnamefont
  {G.}~\bibnamefont {Nishijima}}, \bibinfo {author} {\bibfnamefont
  {K.}~\bibnamefont {Watanabe}}, \bibinfo {author} {\bibfnamefont
  {I.}~\bibnamefont {Sato}}, \bibinfo {author} {\bibfnamefont {H.}~\bibnamefont
  {Kurisita}}, \bibinfo {author} {\bibfnamefont {M.}~\bibnamefont {Narui}},\
  and\ \bibinfo {author} {\bibfnamefont {T.}~\bibnamefont {Shikama}},\
  }\bibfield  {title} {\bibinfo {title} {Framework of collaboration
  investigation on neutron effect on superconducting magnet materials},\ }in\
  \href
  {https://www.nifs.ac.jp/itc/itc18/upload/proceedings_upload/proc_P2-24_Nishimura.pdf}
  {\emph {\bibinfo {booktitle} {Proceedings of the 18th International Toki
  Conference (ITC18)}}}\ (\bibinfo {year} {2008})\ pp.\ \bibinfo {pages}
  {371--374},\ \bibinfo {note} {{P2-24}}\BibitemShut {NoStop}%
\bibitem [{\citenamefont {Kato}\ \emph {et~al.}(1988)\citenamefont {Kato},
  \citenamefont {Watanabe}, \citenamefont {Kazumata}, \citenamefont {Naramoto},
  \citenamefont {Iwata}, \citenamefont {Ikeda}, \citenamefont {Maekawa},\ and\
  \citenamefont {Nakamura}}]{kato88}%
  \BibitemOpen
  \bibfield  {author} {\bibinfo {author} {\bibfnamefont {T.}~\bibnamefont
  {Kato}}, \bibinfo {author} {\bibfnamefont {M.}~\bibnamefont {Watanabe}},
  \bibinfo {author} {\bibfnamefont {Y.}~\bibnamefont {Kazumata}}, \bibinfo
  {author} {\bibfnamefont {H.}~\bibnamefont {Naramoto}}, \bibinfo {author}
  {\bibfnamefont {T.}~\bibnamefont {Iwata}}, \bibinfo {author} {\bibfnamefont
  {Y.}~\bibnamefont {Ikeda}}, \bibinfo {author} {\bibfnamefont
  {H.}~\bibnamefont {Maekawa}},\ and\ \bibinfo {author} {\bibfnamefont
  {T.}~\bibnamefont {Nakamura}},\ }\bibfield  {title} {\bibinfo {title}
  {Radiation effect of {YBa$_2$Cu$_3$O$_{7-y}$} irradiated by $\gamma$-rays and
  14 {MeV} neutrons},\ }\href {https://doi.org/10.1143/JJAP.27.L2097}
  {\bibfield  {journal} {\bibinfo  {journal} {Jpn. J. Appl. Phys.}\ }\textbf
  {\bibinfo {volume} {27}},\ \bibinfo {pages} {L2097} (\bibinfo {year}
  {1988})}\BibitemShut {NoStop}%
\bibitem [{\citenamefont {Pinto}\ \emph {et~al.}(2024)\citenamefont {Pinto},
  \citenamefont {Celentano}, \citenamefont {De~Angelis}, \citenamefont
  {Laviano}, \citenamefont {Masi}, \citenamefont {Pietropaolo}, \citenamefont
  {Tomellini},\ and\ \citenamefont {Torsello}}]{pinto24}%
  \BibitemOpen
  \bibfield  {author} {\bibinfo {author} {\bibfnamefont {V.}~\bibnamefont
  {Pinto}}, \bibinfo {author} {\bibfnamefont {G.}~\bibnamefont {Celentano}},
  \bibinfo {author} {\bibfnamefont {M.}~\bibnamefont {De~Angelis}}, \bibinfo
  {author} {\bibfnamefont {F.}~\bibnamefont {Laviano}}, \bibinfo {author}
  {\bibfnamefont {A.}~\bibnamefont {Masi}}, \bibinfo {author} {\bibfnamefont
  {A.}~\bibnamefont {Pietropaolo}}, \bibinfo {author} {\bibfnamefont
  {M.}~\bibnamefont {Tomellini}},\ and\ \bibinfo {author} {\bibfnamefont
  {D.}~\bibnamefont {Torsello}},\ }\bibfield  {title} {\bibinfo {title} {Effect
  of 14.1 {MeV} fusion neutron irradiation on {YBCO} thin films and commercial
  {REBCO} tapes},\ }\href {https://doi.org/10.1109/TASC.2023.3337057}
  {\bibfield  {journal} {\bibinfo  {journal} {IEEE Trans. Appl. Supercond.}\
  }\textbf {\bibinfo {volume} {34}},\ \bibinfo {pages} {1} (\bibinfo {year}
  {2024})}\BibitemShut {NoStop}%
\bibitem [{\citenamefont {Miralles-Dolz}\ \emph
  {et~al.}(2026{\natexlab{a}})\citenamefont {Miralles-Dolz}, \citenamefont
  {Churchill}, \citenamefont {Panici}, \citenamefont {Schwartz}, \citenamefont
  {Bohm}, \citenamefont {Moreno},\ and\ \citenamefont
  {Wilson}}]{mirallesdolz26}%
  \BibitemOpen
  \bibfield  {author} {\bibinfo {author} {\bibfnamefont {E.}~\bibnamefont
  {Miralles-Dolz}}, \bibinfo {author} {\bibfnamefont {M.}~\bibnamefont
  {Churchill}}, \bibinfo {author} {\bibfnamefont {D.}~\bibnamefont {Panici}},
  \bibinfo {author} {\bibfnamefont {J.}~\bibnamefont {Schwartz}}, \bibinfo
  {author} {\bibfnamefont {T.}~\bibnamefont {Bohm}}, \bibinfo {author}
  {\bibfnamefont {C.}~\bibnamefont {Moreno}},\ and\ \bibinfo {author}
  {\bibfnamefont {P.}~\bibnamefont {Wilson}},\ }\bibfield  {title} {\bibinfo
  {title} {Neutron wall loading in stellarator optimization},\ }\href
  {https://doi.org/10.1088/1361-6587/aeb15b} {\bibfield  {journal} {\bibinfo
  {journal} {Plasma Phys. Control. Fusion}\ }\textbf {\bibinfo {volume} {68}},\
  \bibinfo {pages} {105032} (\bibinfo {year} {2026}{\natexlab{a}})}\BibitemShut
  {NoStop}%
\bibitem [{\citenamefont {Romano}\ \emph {et~al.}(2015)\citenamefont {Romano},
  \citenamefont {Horelik}, \citenamefont {Herman}, \citenamefont {Nelson},
  \citenamefont {Forget},\ and\ \citenamefont {Smith}}]{romano15}%
  \BibitemOpen
  \bibfield  {author} {\bibinfo {author} {\bibfnamefont {P.~K.}\ \bibnamefont
  {Romano}}, \bibinfo {author} {\bibfnamefont {N.~E.}\ \bibnamefont {Horelik}},
  \bibinfo {author} {\bibfnamefont {B.~R.}\ \bibnamefont {Herman}}, \bibinfo
  {author} {\bibfnamefont {A.~G.}\ \bibnamefont {Nelson}}, \bibinfo {author}
  {\bibfnamefont {B.}~\bibnamefont {Forget}},\ and\ \bibinfo {author}
  {\bibfnamefont {K.}~\bibnamefont {Smith}},\ }\bibfield  {title} {\bibinfo
  {title} {{OpenMC}: {A} state-of-the-art {Monte Carlo} code for research and
  development},\ }\href {https://doi.org/10.1016/j.anucene.2014.07.048}
  {\bibfield  {journal} {\bibinfo  {journal} {Ann. Nucl. Energy}\ }\textbf
  {\bibinfo {volume} {82}},\ \bibinfo {pages} {90} (\bibinfo {year}
  {2015})}\BibitemShut {NoStop}%
\bibitem [{\citenamefont {Wilson}\ \emph {et~al.}(2010)\citenamefont {Wilson},
  \citenamefont {Tautges}, \citenamefont {Kraftcheck}, \citenamefont {Smith},\
  and\ \citenamefont {Henderson}}]{wilson10}%
  \BibitemOpen
  \bibfield  {author} {\bibinfo {author} {\bibfnamefont {P.~P.~H.}\
  \bibnamefont {Wilson}}, \bibinfo {author} {\bibfnamefont {T.~J.}\
  \bibnamefont {Tautges}}, \bibinfo {author} {\bibfnamefont {J.~A.}\
  \bibnamefont {Kraftcheck}}, \bibinfo {author} {\bibfnamefont {B.~M.}\
  \bibnamefont {Smith}},\ and\ \bibinfo {author} {\bibfnamefont {D.~L.}\
  \bibnamefont {Henderson}},\ }\bibfield  {title} {\bibinfo {title}
  {Acceleration techniques for the direct use of {CAD}-based geometry in fusion
  neutronics analysis},\ }\href
  {https://doi.org/10.1016/j.fusengdes.2010.05.030} {\bibfield  {journal}
  {\bibinfo  {journal} {Fusion Eng. Des.}\ }\textbf {\bibinfo {volume} {85}},\
  \bibinfo {pages} {1759} (\bibinfo {year} {2010})}\BibitemShut {NoStop}%
\bibitem [{\citenamefont {Landreman}\ \emph {et~al.}(2021)\citenamefont
  {Landreman}, \citenamefont {Medasani}, \citenamefont {Wechsung},
  \citenamefont {Giuliani}, \citenamefont {Jorge},\ and\ \citenamefont
  {Zhu}}]{Landreman2021}%
  \BibitemOpen
  \bibfield  {author} {\bibinfo {author} {\bibfnamefont {M.}~\bibnamefont
  {Landreman}}, \bibinfo {author} {\bibfnamefont {B.}~\bibnamefont {Medasani}},
  \bibinfo {author} {\bibfnamefont {F.}~\bibnamefont {Wechsung}}, \bibinfo
  {author} {\bibfnamefont {A.}~\bibnamefont {Giuliani}}, \bibinfo {author}
  {\bibfnamefont {R.}~\bibnamefont {Jorge}},\ and\ \bibinfo {author}
  {\bibfnamefont {C.}~\bibnamefont {Zhu}},\ }\bibfield  {title} {\bibinfo
  {title} {Simsopt: A flexible framework for stellarator optimization},\ }\href
  {https://doi.org/10.21105/joss.03525} {\bibfield  {journal} {\bibinfo
  {journal} {J. Open Source Softw.}\ }\textbf {\bibinfo {volume} {6}},\
  \bibinfo {pages} {3525} (\bibinfo {year} {2021})}\BibitemShut {NoStop}%
\bibitem [{\citenamefont {Wechsung}\ \emph {et~al.}(2022)\citenamefont
  {Wechsung}, \citenamefont {Landreman}, \citenamefont {Giuliani},
  \citenamefont {Cerfon},\ and\ \citenamefont {Stadler}}]{wechsung22}%
  \BibitemOpen
  \bibfield  {author} {\bibinfo {author} {\bibfnamefont {F.}~\bibnamefont
  {Wechsung}}, \bibinfo {author} {\bibfnamefont {M.}~\bibnamefont {Landreman}},
  \bibinfo {author} {\bibfnamefont {A.}~\bibnamefont {Giuliani}}, \bibinfo
  {author} {\bibfnamefont {A.}~\bibnamefont {Cerfon}},\ and\ \bibinfo {author}
  {\bibfnamefont {G.}~\bibnamefont {Stadler}},\ }\bibfield  {title} {\bibinfo
  {title} {Precise stellarator quasi-symmetry can be achieved with
  electromagnetic coils},\ }\href {https://doi.org/10.1073/pnas.2202084119}
  {\bibfield  {journal} {\bibinfo  {journal} {Proc. Natl. Acad. Sci. U.S.A.}\
  }\textbf {\bibinfo {volume} {119}},\ \bibinfo {pages} {e2202084119} (\bibinfo
  {year} {2022})}\BibitemShut {NoStop}%
\bibitem [{\citenamefont {Hurwitz}\ \emph {et~al.}(2025)\citenamefont
  {Hurwitz}, \citenamefont {Landreman}, \citenamefont {Huslage},\ and\
  \citenamefont {Kaptanoglu}}]{hurwitz25}%
  \BibitemOpen
  \bibfield  {author} {\bibinfo {author} {\bibfnamefont {S.}~\bibnamefont
  {Hurwitz}}, \bibinfo {author} {\bibfnamefont {M.}~\bibnamefont {Landreman}},
  \bibinfo {author} {\bibfnamefont {P.}~\bibnamefont {Huslage}},\ and\ \bibinfo
  {author} {\bibfnamefont {A.~A.}\ \bibnamefont {Kaptanoglu}},\ }\bibfield
  {title} {\bibinfo {title} {Electromagnetic coil optimization for reduced
  {Lorentz} forces},\ }\href {https://doi.org/10.1088/1741-4326/adc9bf}
  {\bibfield  {journal} {\bibinfo  {journal} {Nucl. Fusion}\ }\textbf {\bibinfo
  {volume} {65}},\ \bibinfo {pages} {056044} (\bibinfo {year}
  {2025})}\BibitemShut {NoStop}%
\bibitem [{\citenamefont {Dudt}\ and\ \citenamefont {Kolemen}(2020)}]{dudt20}%
  \BibitemOpen
  \bibfield  {author} {\bibinfo {author} {\bibfnamefont {D.~W.}\ \bibnamefont
  {Dudt}}\ and\ \bibinfo {author} {\bibfnamefont {E.}~\bibnamefont {Kolemen}},\
  }\bibfield  {title} {\bibinfo {title} {{DESC}: {A} stellarator equilibrium
  solver},\ }\href {https://doi.org/10.1063/5.0020743} {\bibfield  {journal}
  {\bibinfo  {journal} {Phys. Plasmas}\ }\textbf {\bibinfo {volume} {27}},\
  \bibinfo {pages} {102513} (\bibinfo {year} {2020})}\BibitemShut {NoStop}%
\bibitem [{\citenamefont {Panici}\ \emph {et~al.}(2023)\citenamefont {Panici},
  \citenamefont {Conlin}, \citenamefont {Dudt}, \citenamefont {Unalmis},\ and\
  \citenamefont {Kolemen}}]{panici23}%
  \BibitemOpen
  \bibfield  {author} {\bibinfo {author} {\bibfnamefont {D.}~\bibnamefont
  {Panici}}, \bibinfo {author} {\bibfnamefont {R.}~\bibnamefont {Conlin}},
  \bibinfo {author} {\bibfnamefont {D.~W.}\ \bibnamefont {Dudt}}, \bibinfo
  {author} {\bibfnamefont {K.}~\bibnamefont {Unalmis}},\ and\ \bibinfo {author}
  {\bibfnamefont {E.}~\bibnamefont {Kolemen}},\ }\bibfield  {title} {\bibinfo
  {title} {The {DESC} stellarator code suite. {Part 1}. {Quick} and accurate
  equilibria computations},\ }\href {https://doi.org/10.1017/S0022377823000272}
  {\bibfield  {journal} {\bibinfo  {journal} {J. Plasma Phys.}\ }\textbf
  {\bibinfo {volume} {89}},\ \bibinfo {pages} {955890303} (\bibinfo {year}
  {2023})}\BibitemShut {NoStop}%
\bibitem [{\citenamefont {Dudt}\ \emph {et~al.}(2023)\citenamefont {Dudt},
  \citenamefont {Conlin}, \citenamefont {Panici},\ and\ \citenamefont
  {Kolemen}}]{dudt23}%
  \BibitemOpen
  \bibfield  {author} {\bibinfo {author} {\bibfnamefont {D.~W.}\ \bibnamefont
  {Dudt}}, \bibinfo {author} {\bibfnamefont {R.}~\bibnamefont {Conlin}},
  \bibinfo {author} {\bibfnamefont {D.}~\bibnamefont {Panici}},\ and\ \bibinfo
  {author} {\bibfnamefont {E.}~\bibnamefont {Kolemen}},\ }\bibfield  {title}
  {\bibinfo {title} {The {DESC} stellarator code suite. {Part} 3.
  {Quasi}-symmetry optimization},\ }\href
  {https://doi.org/10.1017/S0022377823000235} {\bibfield  {journal} {\bibinfo
  {journal} {J. Plasma Phys.}\ }\textbf {\bibinfo {volume} {89}},\ \bibinfo
  {pages} {955890201} (\bibinfo {year} {2023})}\BibitemShut {NoStop}%
\bibitem [{\citenamefont {Nemov}\ \emph {et~al.}(2008)\citenamefont {Nemov},
  \citenamefont {Kasilov}, \citenamefont {Kernbichler},\ and\ \citenamefont
  {Leitold}}]{nemov08}%
  \BibitemOpen
  \bibfield  {author} {\bibinfo {author} {\bibfnamefont {V.~V.}\ \bibnamefont
  {Nemov}}, \bibinfo {author} {\bibfnamefont {S.~V.}\ \bibnamefont {Kasilov}},
  \bibinfo {author} {\bibfnamefont {W.}~\bibnamefont {Kernbichler}},\ and\
  \bibinfo {author} {\bibfnamefont {G.~O.}\ \bibnamefont {Leitold}},\
  }\bibfield  {title} {\bibinfo {title} {Poloidal motion of trapped particle
  orbits in real-space coordinates},\ }\href
  {https://doi.org/10.1063/1.2912456} {\bibfield  {journal} {\bibinfo
  {journal} {Phys. Plasmas}\ }\textbf {\bibinfo {volume} {15}},\ \bibinfo
  {pages} {052501} (\bibinfo {year} {2008})}\BibitemShut {NoStop}%
\bibitem [{\citenamefont {Velasco}\ \emph {et~al.}(2021)\citenamefont
  {Velasco}, \citenamefont {Calvo}, \citenamefont {Mulas}, \citenamefont
  {S{\'a}nchez}, \citenamefont {Parra}, \citenamefont {Cappa},\ and\
  \citenamefont {{the W7-X Team}}}]{velasco21}%
  \BibitemOpen
  \bibfield  {author} {\bibinfo {author} {\bibfnamefont {J.~L.}\ \bibnamefont
  {Velasco}}, \bibinfo {author} {\bibfnamefont {I.}~\bibnamefont {Calvo}},
  \bibinfo {author} {\bibfnamefont {S.}~\bibnamefont {Mulas}}, \bibinfo
  {author} {\bibfnamefont {E.}~\bibnamefont {S{\'a}nchez}}, \bibinfo {author}
  {\bibfnamefont {F.~I.}\ \bibnamefont {Parra}}, \bibinfo {author}
  {\bibfnamefont {{\'A}.}~\bibnamefont {Cappa}},\ and\ \bibinfo {author}
  {\bibnamefont {{the W7-X Team}}},\ }\bibfield  {title} {\bibinfo {title} {A
  model for the fast evaluation of prompt losses of energetic ions in
  stellarators},\ }\href {https://doi.org/10.1088/1741-4326/ac2994} {\bibfield
  {journal} {\bibinfo  {journal} {Nucl. Fusion}\ }\textbf {\bibinfo {volume}
  {61}},\ \bibinfo {pages} {116059} (\bibinfo {year} {2021})}\BibitemShut
  {NoStop}%
\bibitem [{\citenamefont {Kulsrud}\ \emph {et~al.}(1982)\citenamefont
  {Kulsrud}, \citenamefont {Furth}, \citenamefont {Valeo},\ and\ \citenamefont
  {Goldhaber}}]{kulsrud82}%
  \BibitemOpen
  \bibfield  {author} {\bibinfo {author} {\bibfnamefont {R.~M.}\ \bibnamefont
  {Kulsrud}}, \bibinfo {author} {\bibfnamefont {H.~P.}\ \bibnamefont {Furth}},
  \bibinfo {author} {\bibfnamefont {E.~J.}\ \bibnamefont {Valeo}},\ and\
  \bibinfo {author} {\bibfnamefont {M.}~\bibnamefont {Goldhaber}},\ }\bibfield
  {title} {\bibinfo {title} {Fusion reactor plasmas with polarized nuclei},\
  }\href {https://doi.org/10.1103/PhysRevLett.49.1248} {\bibfield  {journal}
  {\bibinfo  {journal} {Phys. Rev. Lett.}\ }\textbf {\bibinfo {volume} {49}},\
  \bibinfo {pages} {1248} (\bibinfo {year} {1982})}\BibitemShut {NoStop}%
\bibitem [{\citenamefont {Parisi}\ \emph {et~al.}(2024)\citenamefont {Parisi},
  \citenamefont {Diallo},\ and\ \citenamefont {Schwartz}}]{parisi24}%
  \BibitemOpen
  \bibfield  {author} {\bibinfo {author} {\bibfnamefont {J.~F.}\ \bibnamefont
  {Parisi}}, \bibinfo {author} {\bibfnamefont {A.}~\bibnamefont {Diallo}},\
  and\ \bibinfo {author} {\bibfnamefont {J.~A.}\ \bibnamefont {Schwartz}},\
  }\bibfield  {title} {\bibinfo {title} {Simultaneous enhancement of tritium
  burn efficiency and fusion power with low-tritium spin-polarized fuel},\
  }\href {https://doi.org/10.1088/1741-4326/ad7da3} {\bibfield  {journal}
  {\bibinfo  {journal} {Nucl. Fusion}\ }\textbf {\bibinfo {volume} {64}},\
  \bibinfo {pages} {126019} (\bibinfo {year} {2024})}\BibitemShut {NoStop}%
\bibitem [{\citenamefont {Bae}\ \emph {et~al.}(2025)\citenamefont {Bae},
  \citenamefont {Borowiec}, \citenamefont {Badalassi}, \citenamefont {Parisi},
  \citenamefont {Diallo}, \citenamefont {Menard}, \citenamefont {Khodak},\ and\
  \citenamefont {Brown}}]{bae25}%
  \BibitemOpen
  \bibfield  {author} {\bibinfo {author} {\bibfnamefont {J.~W.}\ \bibnamefont
  {Bae}}, \bibinfo {author} {\bibfnamefont {K.}~\bibnamefont {Borowiec}},
  \bibinfo {author} {\bibfnamefont {V.}~\bibnamefont {Badalassi}}, \bibinfo
  {author} {\bibfnamefont {J.}~\bibnamefont {Parisi}}, \bibinfo {author}
  {\bibfnamefont {A.}~\bibnamefont {Diallo}}, \bibinfo {author} {\bibfnamefont
  {J.}~\bibnamefont {Menard}}, \bibinfo {author} {\bibfnamefont
  {A.}~\bibnamefont {Khodak}},\ and\ \bibinfo {author} {\bibfnamefont
  {T.}~\bibnamefont {Brown}},\ }\bibfield  {title} {\bibinfo {title}
  {Neutronics analysis of spin-polarized fuel in spherical tokamaks},\ }\href
  {https://doi.org/10.1088/1741-4326/adf3c6} {\bibfield  {journal} {\bibinfo
  {journal} {Nucl. Fusion}\ }\textbf {\bibinfo {volume} {65}},\ \bibinfo
  {pages} {086051} (\bibinfo {year} {2025})}\BibitemShut {NoStop}%
\bibitem [{\citenamefont {Pace}\ \emph {et~al.}(2016)\citenamefont {Pace},
  \citenamefont {Lanctot}, \citenamefont {Jackson}, \citenamefont {Sandorfi},
  \citenamefont {Smith},\ and\ \citenamefont {Wei}}]{pace16}%
  \BibitemOpen
  \bibfield  {author} {\bibinfo {author} {\bibfnamefont {D.~C.}\ \bibnamefont
  {Pace}}, \bibinfo {author} {\bibfnamefont {M.~J.}\ \bibnamefont {Lanctot}},
  \bibinfo {author} {\bibfnamefont {G.~L.}\ \bibnamefont {Jackson}}, \bibinfo
  {author} {\bibfnamefont {A.~M.}\ \bibnamefont {Sandorfi}}, \bibinfo {author}
  {\bibfnamefont {S.~P.}\ \bibnamefont {Smith}},\ and\ \bibinfo {author}
  {\bibfnamefont {X.}~\bibnamefont {Wei}},\ }\bibfield  {title} {\bibinfo
  {title} {Controlling fusion yield in tokamaks with spin polarized fuel, and
  feasibility studies on the {DIII-D} tokamak},\ }\href
  {https://doi.org/10.1007/s10894-015-0015-4} {\bibfield  {journal} {\bibinfo
  {journal} {J. Fusion Energy}\ }\textbf {\bibinfo {volume} {35}},\ \bibinfo
  {pages} {54} (\bibinfo {year} {2016})}\BibitemShut {NoStop}%
\bibitem [{\citenamefont {Baylor}\ \emph {et~al.}(2023)\citenamefont {Baylor},
  \citenamefont {Deur}, \citenamefont {Eidietis}, \citenamefont {Heidbrink},
  \citenamefont {Jackson}, \citenamefont {Liu}, \citenamefont {Lowry},
  \citenamefont {Miller}, \citenamefont {Pace}, \citenamefont {Sandorfi} \emph
  {et~al.}}]{baylor23}%
  \BibitemOpen
  \bibfield  {author} {\bibinfo {author} {\bibfnamefont {L.}~\bibnamefont
  {Baylor}}, \bibinfo {author} {\bibfnamefont {A.}~\bibnamefont {Deur}},
  \bibinfo {author} {\bibfnamefont {N.}~\bibnamefont {Eidietis}}, \bibinfo
  {author} {\bibfnamefont {W.~W.}\ \bibnamefont {Heidbrink}}, \bibinfo {author}
  {\bibfnamefont {G.~L.}\ \bibnamefont {Jackson}}, \bibinfo {author}
  {\bibfnamefont {J.}~\bibnamefont {Liu}}, \bibinfo {author} {\bibfnamefont
  {M.~M.}\ \bibnamefont {Lowry}}, \bibinfo {author} {\bibfnamefont {G.~W.}\
  \bibnamefont {Miller}}, \bibinfo {author} {\bibfnamefont {D.}~\bibnamefont
  {Pace}}, \bibinfo {author} {\bibfnamefont {A.~M.}\ \bibnamefont {Sandorfi}},
  \emph {et~al.},\ }\bibfield  {title} {\bibinfo {title} {Polarized fusion and
  potential in situ tests of fuel polarization survival in a tokamak plasma},\
  }\href {https://doi.org/10.1088/1741-4326/acc3ae} {\bibfield  {journal}
  {\bibinfo  {journal} {Nucl. Fusion}\ }\textbf {\bibinfo {volume} {63}},\
  \bibinfo {pages} {076009} (\bibinfo {year} {2023})}\BibitemShut {NoStop}%
\bibitem [{\citenamefont {Grieger}\ \emph {et~al.}(1992)\citenamefont
  {Grieger}, \citenamefont {Lotz}, \citenamefont {Merkel}, \citenamefont
  {N{\"u}hrenberg}, \citenamefont {Sapper}, \citenamefont {Strumberger},
  \citenamefont {Wobig}, \citenamefont {Burhenn}, \citenamefont {Erckmann},
  \citenamefont {Gasparino}, \citenamefont {Giannone}, \citenamefont
  {Hartfuss}, \citenamefont {Jaenicke}, \citenamefont {K{\"u}hner},
  \citenamefont {Ringler}, \citenamefont {Weller}, \citenamefont {Wagner},
  \citenamefont {{the W7-X Team}},\ and\ \citenamefont {{the W7-AS
  Team}}}]{Grieger1992}%
  \BibitemOpen
  \bibfield  {author} {\bibinfo {author} {\bibfnamefont {G.}~\bibnamefont
  {Grieger}}, \bibinfo {author} {\bibfnamefont {W.}~\bibnamefont {Lotz}},
  \bibinfo {author} {\bibfnamefont {P.}~\bibnamefont {Merkel}}, \bibinfo
  {author} {\bibfnamefont {J.}~\bibnamefont {N{\"u}hrenberg}}, \bibinfo
  {author} {\bibfnamefont {J.}~\bibnamefont {Sapper}}, \bibinfo {author}
  {\bibfnamefont {E.}~\bibnamefont {Strumberger}}, \bibinfo {author}
  {\bibfnamefont {H.}~\bibnamefont {Wobig}}, \bibinfo {author} {\bibfnamefont
  {R.}~\bibnamefont {Burhenn}}, \bibinfo {author} {\bibfnamefont
  {V.}~\bibnamefont {Erckmann}}, \bibinfo {author} {\bibfnamefont
  {U.}~\bibnamefont {Gasparino}}, \bibinfo {author} {\bibfnamefont
  {L.}~\bibnamefont {Giannone}}, \bibinfo {author} {\bibfnamefont {H.~J.}\
  \bibnamefont {Hartfuss}}, \bibinfo {author} {\bibfnamefont {R.}~\bibnamefont
  {Jaenicke}}, \bibinfo {author} {\bibfnamefont {G.}~\bibnamefont
  {K{\"u}hner}}, \bibinfo {author} {\bibfnamefont {H.}~\bibnamefont {Ringler}},
  \bibinfo {author} {\bibfnamefont {A.}~\bibnamefont {Weller}}, \bibinfo
  {author} {\bibfnamefont {F.}~\bibnamefont {Wagner}}, \bibinfo {author}
  {\bibnamefont {{the W7-X Team}}},\ and\ \bibinfo {author} {\bibnamefont {{the
  W7-AS Team}}},\ }\bibfield  {title} {\bibinfo {title} {Physics optimization
  of stellarators},\ }\href {https://doi.org/10.1063/1.860481} {\bibfield
  {journal} {\bibinfo  {journal} {Phys. Fluids B}\ }\textbf {\bibinfo {volume}
  {4}},\ \bibinfo {pages} {2081} (\bibinfo {year} {1992})}\BibitemShut
  {NoStop}%
\bibitem [{\citenamefont {Schilling}(2025)}]{Schilling2025}%
  \BibitemOpen
  \bibfield  {author} {\bibinfo {author} {\bibfnamefont {J.}~\bibnamefont
  {Schilling}},\ }\href {https://arxiv.org/abs/2502.04374} {\bibinfo {title}
  {The numerics of {VMEC++}}} (\bibinfo {year} {2025}),\ \Eprint
  {https://arxiv.org/abs/2502.04374} {arXiv:2502.04374 [physics.plasm-ph]}
  \BibitemShut {NoStop}%
\bibitem [{\citenamefont {Landreman}\ and\ \citenamefont
  {Paul}(2022)}]{landreman22}%
  \BibitemOpen
  \bibfield  {author} {\bibinfo {author} {\bibfnamefont {M.}~\bibnamefont
  {Landreman}}\ and\ \bibinfo {author} {\bibfnamefont {E.}~\bibnamefont
  {Paul}},\ }\bibfield  {title} {\bibinfo {title} {Magnetic fields with precise
  quasisymmetry for plasma confinement},\ }\href
  {https://doi.org/10.1103/physrevlett.128.035001} {\bibfield  {journal}
  {\bibinfo  {journal} {Phys. Rev. Lett.}\ }\textbf {\bibinfo {volume} {128}},\
  \bibinfo {pages} {035001} (\bibinfo {year} {2022})}\BibitemShut {NoStop}%
\bibitem [{\citenamefont {Landreman}(2017)}]{landreman17}%
  \BibitemOpen
  \bibfield  {author} {\bibinfo {author} {\bibfnamefont {M.}~\bibnamefont
  {Landreman}},\ }\bibfield  {title} {\bibinfo {title} {An improved current
  potential method for fast computation of stellarator coil shapes},\ }\href
  {https://doi.org/10.1088/1741-4326/aa57d4} {\bibfield  {journal} {\bibinfo
  {journal} {Nucl. Fusion}\ }\textbf {\bibinfo {volume} {57}},\ \bibinfo
  {pages} {046003} (\bibinfo {year} {2017})}\BibitemShut {NoStop}%
\bibitem [{\citenamefont {Cadena}\ \emph {et~al.}(2025)\citenamefont {Cadena},
  \citenamefont {Merlo}, \citenamefont {Laude}, \citenamefont {Bauer},
  \citenamefont {Agrawal}, \citenamefont {Pascu}, \citenamefont {Savtchouk},
  \citenamefont {Bonauer}, \citenamefont {Guiraud}, \citenamefont {Hudson},\
  and\ \citenamefont {Kaiser}}]{Cadena2025}%
  \BibitemOpen
  \bibfield  {author} {\bibinfo {author} {\bibfnamefont {S.}~\bibnamefont
  {Cadena}}, \bibinfo {author} {\bibfnamefont {A.}~\bibnamefont {Merlo}},
  \bibinfo {author} {\bibfnamefont {E.}~\bibnamefont {Laude}}, \bibinfo
  {author} {\bibfnamefont {A.}~\bibnamefont {Bauer}}, \bibinfo {author}
  {\bibfnamefont {A.}~\bibnamefont {Agrawal}}, \bibinfo {author} {\bibfnamefont
  {M.}~\bibnamefont {Pascu}}, \bibinfo {author} {\bibfnamefont
  {M.}~\bibnamefont {Savtchouk}}, \bibinfo {author} {\bibfnamefont
  {L.}~\bibnamefont {Bonauer}}, \bibinfo {author} {\bibfnamefont
  {E.}~\bibnamefont {Guiraud}}, \bibinfo {author} {\bibfnamefont
  {S.}~\bibnamefont {Hudson}},\ and\ \bibinfo {author} {\bibfnamefont
  {M.}~\bibnamefont {Kaiser}},\ }\bibfield  {title} {\bibinfo {title}
  {Constellaration: A dataset of qi-like stellarator plasma boundaries and
  optimization benchmarks},\ }in\ \href {https://doi.org/10.52202/085713-0520}
  {\emph {\bibinfo {booktitle} {Advances in Neural Information Processing
  Systems 38}}},\ \bibinfo {series and number} {NeurIPS 2025}\ (\bibinfo
  {publisher} {Neural Information Processing Systems Foundation, Inc.
  (NeurIPS)},\ \bibinfo {year} {2025})\ pp.\ \bibinfo {pages}
  {17351--17377}\BibitemShut {NoStop}%
\bibitem [{\citenamefont {Ku}\ and\ \citenamefont
  {Boozer}(2010{\natexlab{b}})}]{kuboozer10nf}%
  \BibitemOpen
  \bibfield  {author} {\bibinfo {author} {\bibfnamefont {L.~P.}\ \bibnamefont
  {Ku}}\ and\ \bibinfo {author} {\bibfnamefont {A.~H.}\ \bibnamefont
  {Boozer}},\ }\bibfield  {title} {\bibinfo {title} {Modular coils and plasma
  configurations for quasi-axisymmetric stellarators},\ }\href
  {https://doi.org/10.1088/0029-5515/50/12/125005} {\bibfield  {journal}
  {\bibinfo  {journal} {Nucl. Fusion}\ }\textbf {\bibinfo {volume} {50}},\
  \bibinfo {pages} {125005} (\bibinfo {year} {2010}{\natexlab{b}})}\BibitemShut
  {NoStop}%
\bibitem [{\citenamefont {Bader}\ \emph {et~al.}(2020)\citenamefont {Bader},
  \citenamefont {Faber}, \citenamefont {Schmitt}, \citenamefont {Anderson},
  \citenamefont {Drevlak}, \citenamefont {Duff}, \citenamefont {Frerichs},
  \citenamefont {Hegna}, \citenamefont {Kruger}, \citenamefont {Landreman},
  \citenamefont {McKinney}, \citenamefont {Singh}, \citenamefont {Schroeder},
  \citenamefont {Terry},\ and\ \citenamefont {Ware}}]{bader20}%
  \BibitemOpen
  \bibfield  {author} {\bibinfo {author} {\bibfnamefont {A.}~\bibnamefont
  {Bader}}, \bibinfo {author} {\bibfnamefont {B.~J.}\ \bibnamefont {Faber}},
  \bibinfo {author} {\bibfnamefont {J.~C.}\ \bibnamefont {Schmitt}}, \bibinfo
  {author} {\bibfnamefont {D.~T.}\ \bibnamefont {Anderson}}, \bibinfo {author}
  {\bibfnamefont {M.}~\bibnamefont {Drevlak}}, \bibinfo {author} {\bibfnamefont
  {J.~M.}\ \bibnamefont {Duff}}, \bibinfo {author} {\bibfnamefont
  {H.}~\bibnamefont {Frerichs}}, \bibinfo {author} {\bibfnamefont {C.~C.}\
  \bibnamefont {Hegna}}, \bibinfo {author} {\bibfnamefont {T.~G.}\ \bibnamefont
  {Kruger}}, \bibinfo {author} {\bibfnamefont {M.}~\bibnamefont {Landreman}},
  \bibinfo {author} {\bibfnamefont {I.~J.}\ \bibnamefont {McKinney}}, \bibinfo
  {author} {\bibfnamefont {L.}~\bibnamefont {Singh}}, \bibinfo {author}
  {\bibfnamefont {J.~M.}\ \bibnamefont {Schroeder}}, \bibinfo {author}
  {\bibfnamefont {P.~W.}\ \bibnamefont {Terry}},\ and\ \bibinfo {author}
  {\bibfnamefont {A.~S.}\ \bibnamefont {Ware}},\ }\bibfield  {title} {\bibinfo
  {title} {Advancing the physics basis for quasi-helically symmetric
  stellarators},\ }\href {https://doi.org/10.1017/S0022377820000963} {\bibfield
   {journal} {\bibinfo  {journal} {J. Plasma Phys.}\ }\textbf {\bibinfo
  {volume} {86}},\ \bibinfo {pages} {905860506} (\bibinfo {year}
  {2020})}\BibitemShut {NoStop}%
\bibitem [{\citenamefont {Najmabadi}\ \emph {et~al.}(2008)\citenamefont
  {Najmabadi}, \citenamefont {Raffray}, \citenamefont {Abdel-Khalik},
  \citenamefont {Bromberg}, \citenamefont {Crosatti}, \citenamefont
  {El-Guebaly}, \citenamefont {Garabedian}, \citenamefont {Grossman},
  \citenamefont {Henderson}, \citenamefont {Ibrahim}, \citenamefont {Ihli},
  \citenamefont {Kaiser}, \citenamefont {Kiedrowski}, \citenamefont {Ku},
  \citenamefont {Lyon}, \citenamefont {Maingi}, \citenamefont {Malang},
  \citenamefont {Martin}, \citenamefont {Mau}, \citenamefont {Merrill},
  \citenamefont {Moore}, \citenamefont {Peipert}, \citenamefont {Petti},
  \citenamefont {Sadowski}, \citenamefont {Sawan}, \citenamefont {Schultz},
  \citenamefont {Slaybaugh}, \citenamefont {Slattery}, \citenamefont
  {Sviatoslavsky}, \citenamefont {Turnbull}, \citenamefont {Waganer},
  \citenamefont {Wang}, \citenamefont {Weathers}, \citenamefont {Wilson},
  \citenamefont {Waldrop}, \citenamefont {Yoda},\ and\ \citenamefont
  {Zarnstorff}}]{najmabadi08}%
  \BibitemOpen
  \bibfield  {author} {\bibinfo {author} {\bibfnamefont {F.}~\bibnamefont
  {Najmabadi}}, \bibinfo {author} {\bibfnamefont {A.~R.}\ \bibnamefont
  {Raffray}}, \bibinfo {author} {\bibfnamefont {S.~I.}\ \bibnamefont
  {Abdel-Khalik}}, \bibinfo {author} {\bibfnamefont {L.}~\bibnamefont
  {Bromberg}}, \bibinfo {author} {\bibfnamefont {L.}~\bibnamefont {Crosatti}},
  \bibinfo {author} {\bibfnamefont {L.}~\bibnamefont {El-Guebaly}}, \bibinfo
  {author} {\bibfnamefont {P.~R.}\ \bibnamefont {Garabedian}}, \bibinfo
  {author} {\bibfnamefont {A.~A.}\ \bibnamefont {Grossman}}, \bibinfo {author}
  {\bibfnamefont {D.}~\bibnamefont {Henderson}}, \bibinfo {author}
  {\bibfnamefont {A.}~\bibnamefont {Ibrahim}}, \bibinfo {author} {\bibfnamefont
  {T.}~\bibnamefont {Ihli}}, \bibinfo {author} {\bibfnamefont {T.~B.}\
  \bibnamefont {Kaiser}}, \bibinfo {author} {\bibfnamefont {B.}~\bibnamefont
  {Kiedrowski}}, \bibinfo {author} {\bibfnamefont {L.~P.}\ \bibnamefont {Ku}},
  \bibinfo {author} {\bibfnamefont {J.~F.}\ \bibnamefont {Lyon}}, \bibinfo
  {author} {\bibfnamefont {R.}~\bibnamefont {Maingi}}, \bibinfo {author}
  {\bibfnamefont {S.}~\bibnamefont {Malang}}, \bibinfo {author} {\bibfnamefont
  {C.}~\bibnamefont {Martin}}, \bibinfo {author} {\bibfnamefont {T.~K.}\
  \bibnamefont {Mau}}, \bibinfo {author} {\bibfnamefont {B.}~\bibnamefont
  {Merrill}}, \bibinfo {author} {\bibfnamefont {R.~L.}\ \bibnamefont {Moore}},
  \bibinfo {author} {\bibfnamefont {R.~J.}\ \bibnamefont {Peipert}}, \bibinfo
  {author} {\bibfnamefont {D.~A.}\ \bibnamefont {Petti}}, \bibinfo {author}
  {\bibfnamefont {D.~L.}\ \bibnamefont {Sadowski}}, \bibinfo {author}
  {\bibfnamefont {M.}~\bibnamefont {Sawan}}, \bibinfo {author} {\bibfnamefont
  {J.~H.}\ \bibnamefont {Schultz}}, \bibinfo {author} {\bibfnamefont
  {R.}~\bibnamefont {Slaybaugh}}, \bibinfo {author} {\bibfnamefont {K.~T.}\
  \bibnamefont {Slattery}}, \bibinfo {author} {\bibfnamefont {G.}~\bibnamefont
  {Sviatoslavsky}}, \bibinfo {author} {\bibfnamefont {A.}~\bibnamefont
  {Turnbull}}, \bibinfo {author} {\bibfnamefont {L.~M.}\ \bibnamefont
  {Waganer}}, \bibinfo {author} {\bibfnamefont {X.~R.}\ \bibnamefont {Wang}},
  \bibinfo {author} {\bibfnamefont {J.~B.}\ \bibnamefont {Weathers}}, \bibinfo
  {author} {\bibfnamefont {P.}~\bibnamefont {Wilson}}, \bibinfo {author}
  {\bibfnamefont {J.~C.}\ \bibnamefont {Waldrop}}, \bibinfo {author}
  {\bibfnamefont {M.}~\bibnamefont {Yoda}},\ and\ \bibinfo {author}
  {\bibfnamefont {M.}~\bibnamefont {Zarnstorff}},\ }\bibfield  {title}
  {\bibinfo {title} {The aries-cs compact stellarator fusion power plant},\
  }\href {https://doi.org/10.13182/fst54-655} {\bibfield  {journal} {\bibinfo
  {journal} {Fusion Sci. Technol.}\ }\textbf {\bibinfo {volume} {54}},\
  \bibinfo {pages} {655} (\bibinfo {year} {2008})}\BibitemShut {NoStop}%
\bibitem [{\citenamefont {Zarnstorff}\ \emph {et~al.}(2001)\citenamefont
  {Zarnstorff}, \citenamefont {Berry}, \citenamefont {Brooks}, \citenamefont
  {Fredrickson}, \citenamefont {Fu}, \citenamefont {Hirshman}, \citenamefont
  {Hudson}, \citenamefont {Ku}, \citenamefont {Lazarus}, \citenamefont
  {Mikkelsen}, \citenamefont {Monticello}, \citenamefont {Neilson},
  \citenamefont {Pomphrey}, \citenamefont {Reiman}, \citenamefont {Spong},
  \citenamefont {Strickler}, \citenamefont {Boozer}, \citenamefont {Cooper},
  \citenamefont {Goldston}, \citenamefont {Hatcher}, \citenamefont {Isaev},
  \citenamefont {Kessel}, \citenamefont {Lewandowski}, \citenamefont {Lyon},
  \citenamefont {Merkel}, \citenamefont {Mynick}, \citenamefont {Nelson},
  \citenamefont {Nuehrenberg}, \citenamefont {Redi}, \citenamefont {Reiersen},
  \citenamefont {Rutherford}, \citenamefont {Sanchez}, \citenamefont
  {Schmidt},\ and\ \citenamefont {White}}]{Zarnstorff2001}%
  \BibitemOpen
  \bibfield  {author} {\bibinfo {author} {\bibfnamefont {M.~C.}\ \bibnamefont
  {Zarnstorff}}, \bibinfo {author} {\bibfnamefont {L.~A.}\ \bibnamefont
  {Berry}}, \bibinfo {author} {\bibfnamefont {A.}~\bibnamefont {Brooks}},
  \bibinfo {author} {\bibfnamefont {E.}~\bibnamefont {Fredrickson}}, \bibinfo
  {author} {\bibfnamefont {G.-Y.}\ \bibnamefont {Fu}}, \bibinfo {author}
  {\bibfnamefont {S.}~\bibnamefont {Hirshman}}, \bibinfo {author}
  {\bibfnamefont {S.}~\bibnamefont {Hudson}}, \bibinfo {author} {\bibfnamefont
  {L.-P.}\ \bibnamefont {Ku}}, \bibinfo {author} {\bibfnamefont
  {E.}~\bibnamefont {Lazarus}}, \bibinfo {author} {\bibfnamefont
  {D.}~\bibnamefont {Mikkelsen}}, \bibinfo {author} {\bibfnamefont
  {D.}~\bibnamefont {Monticello}}, \bibinfo {author} {\bibfnamefont {G.~H.}\
  \bibnamefont {Neilson}}, \bibinfo {author} {\bibfnamefont {N.}~\bibnamefont
  {Pomphrey}}, \bibinfo {author} {\bibfnamefont {A.}~\bibnamefont {Reiman}},
  \bibinfo {author} {\bibfnamefont {D.}~\bibnamefont {Spong}}, \bibinfo
  {author} {\bibfnamefont {D.}~\bibnamefont {Strickler}}, \bibinfo {author}
  {\bibfnamefont {A.}~\bibnamefont {Boozer}}, \bibinfo {author} {\bibfnamefont
  {W.~A.}\ \bibnamefont {Cooper}}, \bibinfo {author} {\bibfnamefont
  {R.}~\bibnamefont {Goldston}}, \bibinfo {author} {\bibfnamefont
  {R.}~\bibnamefont {Hatcher}}, \bibinfo {author} {\bibfnamefont
  {M.}~\bibnamefont {Isaev}}, \bibinfo {author} {\bibfnamefont
  {C.}~\bibnamefont {Kessel}}, \bibinfo {author} {\bibfnamefont
  {J.}~\bibnamefont {Lewandowski}}, \bibinfo {author} {\bibfnamefont {J.~F.}\
  \bibnamefont {Lyon}}, \bibinfo {author} {\bibfnamefont {P.}~\bibnamefont
  {Merkel}}, \bibinfo {author} {\bibfnamefont {H.}~\bibnamefont {Mynick}},
  \bibinfo {author} {\bibfnamefont {B.~E.}\ \bibnamefont {Nelson}}, \bibinfo
  {author} {\bibfnamefont {C.}~\bibnamefont {Nuehrenberg}}, \bibinfo {author}
  {\bibfnamefont {M.}~\bibnamefont {Redi}}, \bibinfo {author} {\bibfnamefont
  {W.}~\bibnamefont {Reiersen}}, \bibinfo {author} {\bibfnamefont
  {P.}~\bibnamefont {Rutherford}}, \bibinfo {author} {\bibfnamefont
  {R.}~\bibnamefont {Sanchez}}, \bibinfo {author} {\bibfnamefont
  {J.}~\bibnamefont {Schmidt}},\ and\ \bibinfo {author} {\bibfnamefont {R.~B.}\
  \bibnamefont {White}},\ }\bibfield  {title} {\bibinfo {title} {Physics of the
  compact advanced stellarator ncsx},\ }\href
  {https://doi.org/10.1088/0741-3335/43/12a/318} {\bibfield  {journal}
  {\bibinfo  {journal} {Plasma Phys. Control. Fusion}\ }\textbf {\bibinfo
  {volume} {43}},\ \bibinfo {pages} {A237} (\bibinfo {year}
  {2001})}\BibitemShut {NoStop}%
\bibitem [{\citenamefont {Rutkowski}\ \emph {et~al.}(2025)\citenamefont
  {Rutkowski}, \citenamefont {Harter},\ and\ \citenamefont
  {Parisi}}]{Rutkowski2025}%
  \BibitemOpen
  \bibfield  {author} {\bibinfo {author} {\bibfnamefont {A.}~\bibnamefont
  {Rutkowski}}, \bibinfo {author} {\bibfnamefont {J.}~\bibnamefont {Harter}},\
  and\ \bibinfo {author} {\bibfnamefont {J.}~\bibnamefont {Parisi}},\
  }\href@noop {} {\bibinfo {title} {Scalable chrysopoeia via $(n,2n)$ reactions
  driven by deuterium-tritium fusion neutrons}} (\bibinfo {year} {2025}),\
  \Eprint {https://arxiv.org/abs/2507.13461} {arXiv:2507.13461} \BibitemShut
  {NoStop}%
\bibitem [{\citenamefont {Molodyk}\ and\ \citenamefont
  {Larbalestier}(2023)}]{molodyk23}%
  \BibitemOpen
  \bibfield  {author} {\bibinfo {author} {\bibfnamefont {A.}~\bibnamefont
  {Molodyk}}\ and\ \bibinfo {author} {\bibfnamefont {D.~C.}\ \bibnamefont
  {Larbalestier}},\ }\bibfield  {title} {\bibinfo {title} {The prospects of
  high-temperature superconductors},\ }\href
  {https://doi.org/10.1126/science.abq4137} {\bibfield  {journal} {\bibinfo
  {journal} {Science}\ }\textbf {\bibinfo {volume} {380}},\ \bibinfo {pages}
  {1220} (\bibinfo {year} {2023})}\BibitemShut {NoStop}%
\bibitem [{\citenamefont {Nemov}\ \emph {et~al.}(1999)\citenamefont {Nemov},
  \citenamefont {Kasilov}, \citenamefont {Kernbichler},\ and\ \citenamefont
  {Heyn}}]{nemov99}%
  \BibitemOpen
  \bibfield  {author} {\bibinfo {author} {\bibfnamefont {V.~V.}\ \bibnamefont
  {Nemov}}, \bibinfo {author} {\bibfnamefont {S.~V.}\ \bibnamefont {Kasilov}},
  \bibinfo {author} {\bibfnamefont {W.}~\bibnamefont {Kernbichler}},\ and\
  \bibinfo {author} {\bibfnamefont {M.~F.}\ \bibnamefont {Heyn}},\ }\bibfield
  {title} {\bibinfo {title} {Evaluation of $1/\nu$ neoclassical transport in
  stellarators},\ }\href {https://doi.org/10.1063/1.873749} {\bibfield
  {journal} {\bibinfo  {journal} {Phys. Plasmas}\ }\textbf {\bibinfo {volume}
  {6}},\ \bibinfo {pages} {4622} (\bibinfo {year} {1999})}\BibitemShut
  {NoStop}%
\bibitem [{\citenamefont {Beidler}\ \emph {et~al.}(2021)\citenamefont
  {Beidler}, \citenamefont {Smith}, \citenamefont {Alonso} \emph
  {et~al.}}]{beidler21}%
  \BibitemOpen
  \bibfield  {author} {\bibinfo {author} {\bibfnamefont {C.~D.}\ \bibnamefont
  {Beidler}}, \bibinfo {author} {\bibfnamefont {H.~M.}\ \bibnamefont {Smith}},
  \bibinfo {author} {\bibfnamefont {A.}~\bibnamefont {Alonso}}, \emph
  {et~al.},\ }\bibfield  {title} {\bibinfo {title} {Demonstration of reduced
  neoclassical energy transport in {W}endelstein 7-{X}},\ }\href
  {https://doi.org/10.1038/s41586-021-03687-w} {\bibfield  {journal} {\bibinfo
  {journal} {Nature}\ }\textbf {\bibinfo {volume} {596}},\ \bibinfo {pages}
  {221} (\bibinfo {year} {2021})}\BibitemShut {NoStop}%
\bibitem [{\citenamefont {Mynick}\ \emph {et~al.}(2010)\citenamefont {Mynick},
  \citenamefont {Pomphrey},\ and\ \citenamefont {Xanthopoulos}}]{mynick10}%
  \BibitemOpen
  \bibfield  {author} {\bibinfo {author} {\bibfnamefont {H.~E.}\ \bibnamefont
  {Mynick}}, \bibinfo {author} {\bibfnamefont {N.}~\bibnamefont {Pomphrey}},\
  and\ \bibinfo {author} {\bibfnamefont {P.}~\bibnamefont {Xanthopoulos}},\
  }\bibfield  {title} {\bibinfo {title} {Optimizing stellarators for turbulent
  transport},\ }\href {https://doi.org/10.1103/PhysRevLett.105.095004}
  {\bibfield  {journal} {\bibinfo  {journal} {Phys. Rev. Lett.}\ }\textbf
  {\bibinfo {volume} {105}},\ \bibinfo {pages} {095004} (\bibinfo {year}
  {2010})}\BibitemShut {NoStop}%
\bibitem [{\citenamefont {Schwab}(1993)}]{schwab93}%
  \BibitemOpen
  \bibfield  {author} {\bibinfo {author} {\bibfnamefont {C.}~\bibnamefont
  {Schwab}},\ }\bibfield  {title} {\bibinfo {title} {Ideal
  magnetohydrodynamics: {G}lobal mode analysis of three-dimensional plasma
  configurations},\ }\href {https://doi.org/10.1063/1.860656} {\bibfield
  {journal} {\bibinfo  {journal} {Phys. Fluids B}\ }\textbf {\bibinfo {volume}
  {5}},\ \bibinfo {pages} {3195} (\bibinfo {year} {1993})}\BibitemShut
  {NoStop}%
\bibitem [{\citenamefont {Lion}\ \emph {et~al.}(2022)\citenamefont {Lion},
  \citenamefont {Warmer},\ and\ \citenamefont {Wang}}]{lion22}%
  \BibitemOpen
  \bibfield  {author} {\bibinfo {author} {\bibfnamefont {J.}~\bibnamefont
  {Lion}}, \bibinfo {author} {\bibfnamefont {F.}~\bibnamefont {Warmer}},\ and\
  \bibinfo {author} {\bibfnamefont {H.}~\bibnamefont {Wang}},\ }\bibfield
  {title} {\bibinfo {title} {A deterministic method for the fast evaluation and
  optimisation of the {3D} neutron wall load for generic stellarator
  configurations},\ }\href {https://doi.org/10.1088/1741-4326/ac6a67}
  {\bibfield  {journal} {\bibinfo  {journal} {Nucl. Fusion}\ }\textbf {\bibinfo
  {volume} {62}},\ \bibinfo {pages} {076040} (\bibinfo {year}
  {2022})}\BibitemShut {NoStop}%
\bibitem [{\citenamefont {Miralles-Dolz}\ \emph
  {et~al.}(2026{\natexlab{b}})\citenamefont {Miralles-Dolz}, \citenamefont
  {Churchill}, \citenamefont {Schwartz}, \citenamefont {Hidalgo-Salaverri},
  \citenamefont {Moreno}, \citenamefont {Bohm},\ and\ \citenamefont
  {Wilson}}]{mirallesdolz26sr}%
  \BibitemOpen
  \bibfield  {author} {\bibinfo {author} {\bibfnamefont {E.}~\bibnamefont
  {Miralles-Dolz}}, \bibinfo {author} {\bibfnamefont {M.}~\bibnamefont
  {Churchill}}, \bibinfo {author} {\bibfnamefont {J.}~\bibnamefont {Schwartz}},
  \bibinfo {author} {\bibfnamefont {J.}~\bibnamefont {Hidalgo-Salaverri}},
  \bibinfo {author} {\bibfnamefont {C.}~\bibnamefont {Moreno}}, \bibinfo
  {author} {\bibfnamefont {T.}~\bibnamefont {Bohm}},\ and\ \bibinfo {author}
  {\bibfnamefont {P.}~\bibnamefont {Wilson}},\ }\bibfield  {title} {\bibinfo
  {title} {Stellarator design exploration using symbolic-regression neutronics
  surrogates},\ }\href {https://doi.org/10.1109/TPS.2026.3680528} {\bibfield
  {journal} {\bibinfo  {journal} {IEEE Trans. Plasma Sci.}\ }\textbf {\bibinfo
  {volume} {54}},\ \bibinfo {pages} {2879} (\bibinfo {year}
  {2026}{\natexlab{b}})}\BibitemShut {NoStop}%
\bibitem [{\citenamefont {Heidbrink}\ \emph {et~al.}(2024)\citenamefont
  {Heidbrink}, \citenamefont {Baylor}, \citenamefont {B{\"u}scher},
  \citenamefont {Engels}, \citenamefont {Garcia}, \citenamefont {Ghiozzi},
  \citenamefont {Miller}, \citenamefont {Sandorfi}, \citenamefont {Wei},\ and\
  \citenamefont {Zheng}}]{heidbrink24}%
  \BibitemOpen
  \bibfield  {author} {\bibinfo {author} {\bibfnamefont {W.~W.}\ \bibnamefont
  {Heidbrink}}, \bibinfo {author} {\bibfnamefont {L.~R.}\ \bibnamefont
  {Baylor}}, \bibinfo {author} {\bibfnamefont {M.}~\bibnamefont {B{\"u}scher}},
  \bibinfo {author} {\bibfnamefont {R.~W.}\ \bibnamefont {Engels}}, \bibinfo
  {author} {\bibfnamefont {A.~V.}\ \bibnamefont {Garcia}}, \bibinfo {author}
  {\bibfnamefont {A.~G.}\ \bibnamefont {Ghiozzi}}, \bibinfo {author}
  {\bibfnamefont {G.~W.}\ \bibnamefont {Miller}}, \bibinfo {author}
  {\bibfnamefont {A.~M.}\ \bibnamefont {Sandorfi}}, \bibinfo {author}
  {\bibfnamefont {X.}~\bibnamefont {Wei}},\ and\ \bibinfo {author}
  {\bibfnamefont {X.}~\bibnamefont {Zheng}},\ }\bibfield  {title} {\bibinfo
  {title} {A research program to measure the lifetime of spin polarized fuel},\
  }\href {https://doi.org/10.3389/fphy.2024.1355212} {\bibfield  {journal}
  {\bibinfo  {journal} {Front. Phys.}\ }\textbf {\bibinfo {volume} {12}},\
  \bibinfo {pages} {1355212} (\bibinfo {year} {2024})}\BibitemShut {NoStop}%
\bibitem [{\citenamefont {Risse}\ \emph {et~al.}(2017)\citenamefont {Risse},
  \citenamefont {Bykov}, \citenamefont {Nagel}, \citenamefont {Rummel},
  \citenamefont {Bosch}, \citenamefont {Carls}, \citenamefont {M{\"o}nnich},\
  and\ \citenamefont {Schneider}}]{risse17}%
  \BibitemOpen
  \bibfield  {author} {\bibinfo {author} {\bibfnamefont {K.}~\bibnamefont
  {Risse}}, \bibinfo {author} {\bibfnamefont {V.}~\bibnamefont {Bykov}},
  \bibinfo {author} {\bibfnamefont {M.}~\bibnamefont {Nagel}}, \bibinfo
  {author} {\bibfnamefont {T.}~\bibnamefont {Rummel}}, \bibinfo {author}
  {\bibfnamefont {H.-S.}\ \bibnamefont {Bosch}}, \bibinfo {author}
  {\bibfnamefont {A.}~\bibnamefont {Carls}}, \bibinfo {author} {\bibfnamefont
  {T.}~\bibnamefont {M{\"o}nnich}},\ and\ \bibinfo {author} {\bibfnamefont
  {M.}~\bibnamefont {Schneider}},\ }\bibfield  {title} {\bibinfo {title} {First
  operational phase of the superconducting magnet system of {W}endelstein
  7-{X}},\ }\href {https://doi.org/10.1016/j.fusengdes.2017.03.050} {\bibfield
  {journal} {\bibinfo  {journal} {Fusion Eng. Des.}\ }\textbf {\bibinfo
  {volume} {124}},\ \bibinfo {pages} {10} (\bibinfo {year} {2017})}\BibitemShut
  {NoStop}%
\bibitem [{\citenamefont {Sarasola}\ \emph {et~al.}(2020)\citenamefont
  {Sarasola}, \citenamefont {Wesche}, \citenamefont {Ivashov}, \citenamefont
  {Sedlak}, \citenamefont {Uglietti},\ and\ \citenamefont
  {Bruzzone}}]{sarasola20}%
  \BibitemOpen
  \bibfield  {author} {\bibinfo {author} {\bibfnamefont {X.}~\bibnamefont
  {Sarasola}}, \bibinfo {author} {\bibfnamefont {R.}~\bibnamefont {Wesche}},
  \bibinfo {author} {\bibfnamefont {I.}~\bibnamefont {Ivashov}}, \bibinfo
  {author} {\bibfnamefont {K.}~\bibnamefont {Sedlak}}, \bibinfo {author}
  {\bibfnamefont {D.}~\bibnamefont {Uglietti}},\ and\ \bibinfo {author}
  {\bibfnamefont {P.}~\bibnamefont {Bruzzone}},\ }\bibfield  {title} {\bibinfo
  {title} {Progress in the design of a hybrid {HTS}-{Nb$_3$Sn}-{NbTi} central
  solenoid for the {EU DEMO}},\ }\href
  {https://doi.org/10.1109/TASC.2020.2965066} {\bibfield  {journal} {\bibinfo
  {journal} {IEEE Trans. Appl. Supercond.}\ }\textbf {\bibinfo {volume} {30}},\
  \bibinfo {pages} {1} (\bibinfo {year} {2020})}\BibitemShut {NoStop}%
\end{thebibliography}%
\end{document}